Research Report

# CALCULATION OF FORENSIC LIKELIHOOD RATIOS: USE OF MONTE CARLO SIMULATIONS TO COMPARE THE OUTPUT OF SCORE-BASED APPROACHES WITH TRUE LIKELIHOOD-RATIO VALUES

**Geoffrey Stewart Morrison**[1,2]

[1]School of Electrical Engineering & Telecommunications,

University of New South Wales, UNSW Sydney, NSW 2052, Australia

[2]Department of Linguistics, University of Alberta, Edmonton, Alberta, T6G 2E7, Canada

geoff-morrison@forensic-evaluation.net

## Abstract

A group of approaches for calculating forensic likelihood ratios first calculates scores which quantify the degree of difference or the degree of similarity between pairs of samples, then converts those scores to likelihood ratios. In order for a score-based approach to produce a forensically interpretable likelihood ratio, however, in addition to accounting for the similarity of the questioned sample with respect to the known sample, it must also account for the typicality of the questioned sample with respect to the relevant population. The present paper explores a number of score-based approaches using different types of scores and different procedures for converting scores to likelihood ratios. Monte Carlo simulations are used to compare the output of these approaches to true likelihood-ratio values calculated on the basis of the distribution specified for a simulated population. The inadequacy of approaches based on similarity-only or difference-only scores is illustrated, and the relative performance of different approaches which take account of both similarity and typicality is assessed.

The first draft of the present paper was written in 2013. It has subsequently undergone several revisions. The data of the current version appears above. A version was presented at the 9th International Conference on Forensic Inference and Statistics, Leiden, The Netherlands, 19–22 August 2014. An abridged version has been submitted for consideration for publication in a refereed journal.

# 1  INTRODUCTION

In order to calculate a forensic likelihood ratio one must assess the probability of the evidence given the prosecution hypothesis, $p(E \mid H_\mathrm{p})$, versus the probability of the evidence given the defence hypothesis, $p(E \mid H_\mathrm{d})$, as in Eq. 1a.[1] The present paper is concerned with data from types of evidence such as glass fragments, fibres, fingerprints, voice recordings, etc. for which each fragment, fibre, mark, recording, etc. (hereafter "token") is a sample of a pane of glass, a garment, a finger, a speaker's voice, etc. (hereafter "source"), and for which measurements made on the tokens are continuously valued and have (across-token or across-measurement) within-source variability due to intrinsic variability within the source, or variability due to the transfer mechanism whereby a token becomes available for analysis, or variability in the measurement procedure itself. For example: voice recordings presented for forensic comparison are usually of different things being said and even if a speaker attempts to say exactly the same thing exactly the same way twice the acoustic signals would not be expected to be identical; finger marks from the same finger can vary due to different patterns of skin distortion on different surfaces; repeated measures of refractive index on the same glass fragment may have different values depending on the precision of the measurement device. For concreteness and simplicity, we will adopt a common prosecution hypothesis that the suspect is the origin of the token of questioned identity (hereafter "offender token") versus a common defence hypothesis that some other source randomly drawn form the relevant population is the origin of the offender token (Eq 1b). This allows for a terminological convenience, which is not a necessary part of the argumentation but simplifies its expression: the numerator of the likelihood ratio can be considered a quantification of the similarity of the offender token with respect to the suspect, and the denominator can be considered a quantification of the typicality of the offender token with respect to the relevant population (Eq 1c).

---

[1]There are many introduction to the likelihood-ratio framework for the evaluation of forensic evidence, including Robertson & Vignaux (1995), Balding (2005), Aitken et al. (2010), Morrison (2010) and the present paper assumes familiarity with the framework.

$$LR = \frac{p\left(E|H_{\mathrm{p}}\right)}{p\left(E|H_{\mathrm{d}}\right)} \tag{1a}$$

$$LR = \frac{\text{probability of the measured properties of the offender token}}{\text{had it come from the suspect}} \atop {\text{probability of the measured properties of the offender token} \atop \text{had it come from some other source randomly drawn the relevant population}} \tag{1b}$$

$$LR = \frac{\textbf{similarity} \text{ of the offender token with respect to the suspect}}{\textbf{typicality} \text{ of the offender token with respect to the relevant population}} \tag{1c}$$

Score-based approaches are popular for calculating forensic likelihood ratios (Hepler et al., 2012; Abraham et al., 2013). A score is a measurement which quantifies the degree of similarity or difference between pairs of samples, where each member of the pair consists of one or more tokens. A score may also take account of typicality with respect to a sample of the relevant population. Training scores are then used to train models which are used to convert new scores to likelihood ratios. Training scores are calculated for numerous pairs of tokens drawn from a database representative of the relevant population. Some pairs are known to come from the same source as each other and other pairs are known to come from different sources from each other. One may then build a model of the distribution of the values of scores known to come from same-origin pairs and another model of the distribution of the values of scores known to come from different-origin pairs. One then evaluates the probability density of the same-origin model at the value of the score derived from comparison of the suspect and offender tokens – this is used as the numerator of the likelihood ratio. One also evaluates the probability density of the different-origin model at the value of the score derived from comparison of the suspect and offender tokens – this is used as the denominator of the likelihood ratio, see Eq. 2:

$$LR_{q,k} = f(S_{q,k} \mid \mathcal{M}_{\mathrm{so}}) \, / \, f(S_{q,k} \mid \mathcal{M}_{\mathrm{do}}), \tag{2a}$$

$$f(S_{q,k} \mid \mathcal{M}_{\mathrm{so}}) = f(\, S_{q,k} \mid S_{i,j}, \, i = j = \{1..n\} \,), \tag{2b}$$

$$f(S_{q,k} \mid \mathcal{M}_{\mathrm{do}}) = f(\, S_{q,k} \mid S_{i,j}, \, i \neq j, \, i = \{1..n-1\}, \, j = \{i+1..n\} \,), \tag{2c}$$

$$S_{q,k} = g(x_q, \boldsymbol{x}_k), \qquad\qquad S_{i,j} = g(x_i, \boldsymbol{x}_j), \tag{2d}$$

or

$$S_{q,k} = g(x_q, \boldsymbol{x}_k, \boldsymbol{B}), \qquad\qquad S_{i,j} = g(x_i, \boldsymbol{x}_j, \boldsymbol{B}), \tag{2e}$$

where $S_{q,k}$ and $LR_{q,k}$ are respectively the score and the likelihood ratio calculated for the comparison of offender token $x_q$ (questioned identity) and suspect tokens $\boldsymbol{x}_k$ (known identity).[2] $f(S|\mathcal{M})$ is the evaluation of a probability-density-function model $\mathcal{M}$ at a score value $S$. $\mathcal{M}_{\mathrm{so}}$ and $\mathcal{M}_{\mathrm{do}}$ are probability-density-function models trained using a training set of same-origin and different-origin scores respectively. A score $S_{i,j}$ is a value based on a function $g(x_i, \boldsymbol{x}_j)$ which quantifies the degree of similarity or the degree of difference between a token $x_i$ sampled from source $i$ and tokens $\boldsymbol{x}_j$ sampled from source $j$, or based on a function $g(x_i, \boldsymbol{x}_j, \boldsymbol{B})$ which also takes account of the typicality of $x_i$ with respect to a sample of the relevant population $\boldsymbol{B}$ (a "background sample"). Note that for the same-origin model $i = j$ (Eq. 2b), and for the different-origin model $i \neq j$ (Eq. 2c). When $i = j$, $x_i$ and $\boldsymbol{x}_j$ are tokens sampled from the same source but token $x_i$ is not the same token as any of the tokens in $\boldsymbol{x}_j$. When $i \neq j$, $x_i$ and $\boldsymbol{x}_j$ could be tokens drawn from separate data sets or drawn from the same set in a manner such as suggested by $i = \{1..n-1\}$, $j = \{i+1..n\}$ in Eq. 2c. Scores can also be "anchored", e.g., the same-origin model can be suspect anchored by training it using scores derived from comparison of the suspect tokens with a second set of suspect tokens (a suspect control database), and the different-origin model can be offender-anchored by training it using scores derived from comparisons of the offender token with tokens from different sources in a background sample. Non-anchored scores are calculated using same-origin and different origin-pairs in a background database, without direct involvement of either the suspect or offender tokens.

---

[2]Note that in the present paper we assume there is only one measurement made on one token from the offender. It would be possible to make multiple measurements on a single token. In some circumstances an average of those measurements could be used in the calculation of a likelihood ratio, but this would not be appropriate if the distribution of these measurements is multimodal, e.g., for spectral measurements made every few milliseconds throughout a voice recording. In this circumstance, a likelihood ratio could be calculated for each individual measurement, then these averaged and treated as a score to be calibrated (see Morrison, 2013). A true likelihood-ratio value could be calculated with respect to each individual measurement made on the offender token, but it is not clear what would constitute a true likelihood ratio with respect to the combination of all of these measurements.

The present paper proceeds by first describing a Monte Carlo simulation and the calculation of a set of true likelihood ratios based on the distribution of the population specified for the simulation. Several procedures for calculating likelihood ratios are then described, sets of likelihood ratios are calculated using samples drawn from the simulated population, and these are compared with the true likelihood ratios. Hepler et al. (2012) compared the output of score-based approaches with true likelihood-ratio values from simple specified population distributions for which there are analytic solutions. Monte Carlo simulation provides a numerical solution which can potentially be extended to more complex distributions, as illustrated in Appendix A.[3]

We begin with a direct method for the calculation of likelihood ratios, then examine several score-based methods: difference-only scores (non anchored), similarity-only scores (non anchored), similarity-and-typicality scores (non anchored), similarity scores for the numerator and typicality scores for the denominator (suspect-anchored scores for the numerator and offender-anchored scores for the denominator), suspect-anchored scores for both the numerator and denominator, and support-vector-machine scores. We also explore the combination of the similarity-and-typicality scores with a number of procedures for converting scores to likelihood ratios: kernel density, logistic regression, equal-variance Gaussian, separate-variance Gaussian, and pool adjacent violators. Some of these are found to produce better results than others, which in part can be predicted on the basis of modelling assumptions and data requirements. Methods based on difference-only scores or similarity-only scores are problematic from a theoretical perspective, and empirically they are also found to perform poorly. Methods based on similarity scores for the numerator and typicality scores for the denominator, methods based on suspect-anchored scores in both the numerator and denominator, and methods based on support-vector-machine scores are empirically found to perform relatively poorly on the tests conducted in the present study. In contrast, when combined with appropriate score-to-likelihood-ratio-conversion procedures, methods based on scores which take account of both similarity and typicality are shown to be able to produce output which may be considered reasonable approximations of the true likelihood-ratio values.

---

[3] Ali, Spreeuwers, & Veldhuis (2012) (see also Ali, 2014) used Monte Carlo simulations to compare different methods for converting scores to likelihood ratios (the methods included kernel density, logistic regression, and pool adjacent violators), but their starting point was a distribution of scores not the distribution of features in the original feature space, hence they did not have true likelihood-ratio values against which to compare the output.

## 2   MONTE CARLO SIMULATION

Except in trivial cases, in real life one cannot know the true statistical distribution for a population. In real life one cannot, therefore, know the true value of the probability density function for the population given the evidence. One cannot, therefore, know the true value of a likelihood ratio. In a test scenario, one can compare the value of a likelihood ratio with knowledge as to the truth of whether it was derived from a same-origin or a different-origin comparison – all else being equal one would prefer larger likelihood-ratio values from comparisons one knows to be same-origin comparisons and smaller likelihood-ratio values from comparisons one knows to be different-origin comparisons (see Morrison, 2011). One cannot, however, know the true value for the likelihood ratio in the sense of knowing the ratio of the true likelihood of the population distribution for the prosecution hypothesis (the true underlying distribution for the suspect) and of the population distribution for the defence hypothesis (the true underlying distribution for the relevant population). Note that the latter definition of a true likelihood-ratio value is predicated on knowing the true population distributions for a particular kind of property. A true likelihood-ratio value predicated on knowing the true population distributions for the property of length would not be expected to the same as a true likelihood-ratio value predicated on knowing the true population distribution for the property of width.

In real life, since one does not know the true population distributions, one trains models on data sampled from the populations, and the models are used as estimates of the population distributions (see discussion in Morrison & Stoel, 2014). In contrast to real life, Monte Carlo simulation allows one to specify the probability distributions of simulated populations. Since one has specified the distribution for the simulated populations, one knows their true distributions, and one can therefore calculate true likelihood values with respect to these populations. If the populations are specified in terms of particular probability models and their parameter values, one can then use a pseudo-random number generator to generate a any desired number of sets of samples from the specified populations, and those sample sets can be of any desired size. One can then use different models to calculate likelihood ratios on the basis of these sample sets and compare the model outputs with the true likelihood-ratio values for the simulated population.

The tests reported below are based on the distributions shown in Fig. 1a. The plot of the distribution of the simulated population for the defence hypothesis is marked "relevant population". Two different population distributions for two different prosecution hypotheses are each marked "suspect". For simplicity these will henceforth be referred to as the "relevant population" and the "suspect" distributions. The distributions for the simulated suspects are univariate normal distributions one with a mean of 0 and the other with a mean of 2, and each with a within-source standard deviation of 1. The simulated population distribution for the defence hypothesis is based on a univariate normal distribution with a grand mean of 0 and a between-source standard deviation of 2. A pseudo-random-number generator was used to generate 1000 points given this distribution. Each of these points is the initial mean value for a source in a finite population of 1000 sources. The distribution of tokens within each source is initially specified as a normal distribution with a within-source standard deviation of 1.[4] For each source the pseudo-random-number generator was used to generate 10 tokens on the basis of its initial mean and standard deviation. The resulting tokens were then used to calculate a revised mean and standard deviation for each source, and these values were saved as the parameters for the simulated population. Because of the last step, the within-source standard deviation in the synthetic population is not equal across sources.

The population and the samples were generated in MATLAB (Mathworks, 2013), and MATLAB's random-number stream was first reset so that exactly the same simulated data are generated every time the code is run. A copy of the MATLAB code running all the demonstrations discussed in the present paper is available from http://geoff-morrison.net/#ICFIS2014.

---

[4]The overall resulting distribution is therefore the sum of two normal distributions, so is itself a normal distribution with an expected grand mean of $\mu_1 + \mu_2 = 0 + 0 = 0$ and expected standard deviation of $\sqrt{(\sigma_1{}^2 + \sigma_2{}^2)} = \sqrt{(2^2 + 1^2)} = \sqrt{5} \approx 2.24$.

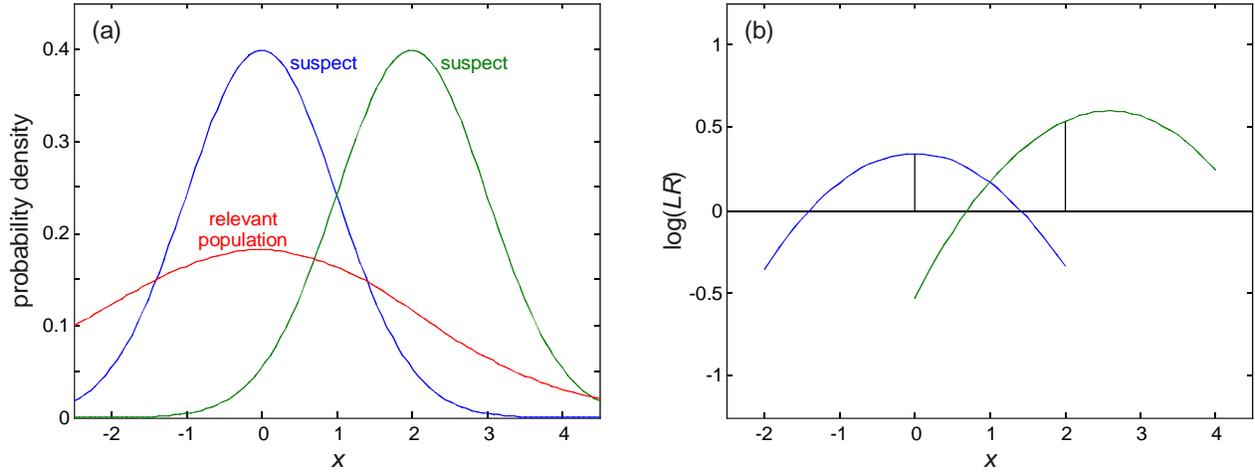

**Figure 1.** (a) Distribution of the simulated relevant population, plus two simulated suspects. (b) True likelihood-ratio values derived from the simulated relevant population and specified values for pairs of suspect means and offender tokens. The vertical lines indicate the values of the suspect means from the same simulated suspects as in Fig. 1a.

## 3  TRUE LIKELIHOOD-RATIO VALUES

A set of likelihood-ratio values were calculated given the simulated relevant population and suspect distributions. The calculated likelihood-ratio values are true likelihood-ratio values with respect to the population distributions specified for the simulation. Values were calculated for two simulated suspects, one with a mean of 0 and the other with a mean of 2 and each with a standard deviation of 1 (see Fig. 1a). These represent the true underlying distributions for the suspects. Each suspect was compared with a series of simulated offenders, 1 token from each offender. The offender tokens had a range of values from 2 less than the suspect mean to 2 greater than the suspect mean, and were spaced 0.1 apart (one series of 41 offender tokens centred around and compared with the first suspect and another series of 41 offender tokens centred around and compared with the second suspect). A likelihood ratio $LR_{q,k}$ comparing an offender token $x_q$ with its respective suspect $k$ was calculated as in Eq. 3:

$$LR_{q,k} = f(x_q \mid H_p) \,/\, f(x_q \mid H_d), \tag{3a}$$

$$f(x_q \mid H_p) = \varphi(x_q \mid \mu_k, \sigma_k), \tag{3b}$$

$$f\left(x_q \,|\, H_\mathrm{d}\right) = \frac{1}{J}\sum_{j=1}^{J} \varphi\left(x_q \,|\, \mu_j, \sigma_j\right), \tag{3c}$$

$$\varphi\left(x_q \,|\, \mu, \sigma\right) = \frac{1}{\sigma\sqrt{2\pi}}\, e^{\frac{-\left(x_q-\mu\right)^2}{2\sigma^2}}, \tag{3d}$$

where $f(x_q \mid H)$ is the likelihood of hypothesis $H$ given the offender token $x_q$. The prosecution hypothesis $H_\mathrm{p}$ is quantified by a normal distribution $\varphi$ with mean $\mu_k$ corresponding to the suspect mean ($\mu_k$ is specified as either 0 or 2) and standard deviation $\sigma_k$ ($\sigma_k$ is specified as 1 for both $k$). The defence hypothesis $H_\mathrm{d}$ is quantified by an equally-weighted sum of $J$ normal distributions ($J = 1000$) with each distribution having a mean $\mu_j$ and standard deviation $\sigma_j$, the mean and the standard deviation of the $j$th source as specified in the simulated population. For simplicity, Eq. 3d drops the subscripts, which would be either $k$ or $j$.

The resulting true likelihood-ratio values are shown in Fig. 1b. The leftmost curve joints the likelihood-ratio values calculated for the suspect with a mean of 0 and its series of 41 offender tokens ranging from −2 to +2, and the rightmost curve joins the likelihood-ratio values calculated for the suspect with a mean of 2 and its series of 41 offender tokens ranging from 0 to +4. $x$ values in the original feature space are represented on the abscissa, and vertical lines are draw at the points corresponding to the suspect means, $\mu_k$. Log-base-ten-likelihood-ratio values, log($LR$), are represented on the ordinate.

Note that since the first suspect has a mean of 0, which is the same as the mean of the relevant population, the true likelihood-ratio values from comparisons of this suspect and its associated offender tokens are symmetrical (see Fig. 1b) in the sense that likelihood ratios corresponding to offender tokens equidistant but in opposite directions away from the suspect mean have equal likelihood-ratio values.[5] In contrast, because the second suspect has a mean on a slope of the relevant-population distribution (see Fig. 1a), the true likelihood-ratio values from comparisons of this suspect and its associated offender tokens are asymmetrical (see Fig. 1b). This contrast is related to the relative atypicality of the second set of suspect and offender tokens compared to the first set.

---

[5] The symmetry will be approximate: A slight deviation from symmetry will arise because the distribution of the actual simulated population is unlikely to be exactly normally distributed with a grand mean of 0. Such deviation from symmetry is unlikely to be perceptible on the plots.

## 4 DIRECT CALCULATION OF LIKELIHOOD-RATIO VALUES

The likelihood-ratio values calculated above were true likelihood-ratio values calculated given knowledge of the specified population distributions. In real life, sample data are used to train models which are estimates of the population distributions. In the Monte Carlo simulations reported below, sample sets will be drawn pseudo-randomly from the simulated population, selecting 100 of the 1000 sources, then for each selected source using the pseudo-random-number generator to generate 30 sample tokens.

Even if one did not know the true underlying data structure for the simulated population, examination of a reasonable-size set of sample data drawn from the population would suggest that the within-source and between-source distributions are unimodal and that an assumption of normality would be reasonable. The model given in Eq. 3 would therefore be an appropriate way of directly calculating estimates of likelihood-ratio values. To calculate a likelihood ratio on the basis of a sample of the relevant population, rather than using the relevant population's specified source means and within-source standard deviations, each $\mu_j$ would be a sample source mean calculated from 30 sample tokens generated for each source $j$ included in the sample. There are two basic options for calculating the standard deviation, either calculate a pooled standard deviation across all $J = 100$ sample sources, i.e, in Eq. 3 $\sigma_j = \sigma_{\text{pooled}}$ for all $j$ (Eq. 4a), or calculate a separate standard deviation for each source based on its own $T = 30$ tokens only (separate $\sigma_j$ for each $j$, Eq. 4b). Throughout the paper we use unbiassed least-squares estimates of standard deviations. To calculate likelihood ratios for a sample of the population using Eq. 3 we set the offender tokens $x_q$ and suspect means $\mu_k$ at their specified values as before, but we base the standard deviation for a suspect $\sigma_k$ either on a sample of 30 tokens generated using a within-source standard deviation of 1 (a different sample of 30 tokens is used for each suspect) or we use the pooled standard deviation calculated using the background data. Although we have specified a fixed value for each suspect mean $\mu_k$, it represents a mean calculated on the basis of 30 tokens.

$$\sigma_{\text{pooled}} = \sqrt{\frac{1}{J}\sum_{j=1}^{J}\sigma_j^2} \tag{4a}$$

$$\sigma_j = \sqrt{\frac{1}{T-1}\sum_{t=1}^{T}\left(x_{j,t}-\mu_j\right)^2} \tag{4b}$$

Fig. 2 shows plots of likelihood-ratio values calculated using pooled standard-deviation estimates. The solid line in Fig. 2a is calculated on the basis of the first sample set of 100 sources with 30 tokens per source. The three solid lines in Fig. 2b represent the 5th percentile, median, and 95th percentile from 100 sets of likelihood-ratio values calculated on the basis of 100 samples sets of 100 sources each with 30 tokens per source. The median and percentiles are calculated on a per offender basis, not on a per sample-set basis.[6] The dashed lines in both panels represent the true likelihood-ratio values. Note that the output of this procedure has replicated the asymmetric nature of the true likelihood-ratio values. It also has only a small deviation from the true values, i.e., a small error. As a means of quantifying this error, a root-mean-square (RMS) difference between each estimated log-likelihood-ratio value and its corresponding true log-likelihood-ratio value was calculated across all the suspect and offender pairs for each sample set, and the mean and standard deviation of these RMS values was taken across all 100 sample sets. The resulting median RMS error value was 0.050. Fig. 3 provides boxplots of the RMS error values and allows for visual comparison across different approaches.

Fig. 4 shows plots of likelihood-ratio values calculated using the same sample sets as in Fig. 2 but using separate standard-deviation estimates for each source. Note that the output of this procedure has also replicated the asymmetric nature of the true likelihood-ratio values, and on average has resulted in estimates of likelihood-ratio values which are close to the true values, but relative to the pooled-standard-deviation procedure this procedure has produced a greater variance in those estimates. Note that the separate-standard-deviation approach has an extreme outlier (see Fig. 3). The median RMS error was 0.089.

---

[6]This means that the top line on the right in Fig. 2b, for example, is unlikely to represent the results from any single sample set. Individual sample sets may produce relatively high values for some offenders (e.g., for $\mu_k = 2$, $x_q = 2.5$ in Fig. 2a) and relatively low values for other offenders (e.g., for $\mu_k = 2$, $x_q = 4$ in Fig. 2a).

A pooled procedure will always benefit from having a relatively larger amount of data to train a relatively smaller number of parameter estimates, and it thus results in a lower variance model, although a potentially higher bias model. The lower variance may offset the higher bias and lead to better performance. In this case the pooled-standard-deviation model has performed better, even though we know that the within-source standard deviations in the underlying population distribution vary from source to source. For simplicity, wherever applicable, only likelihood-ratios calculated using pooled-standard-deviation estimates will be reported for the scoring methods described below.

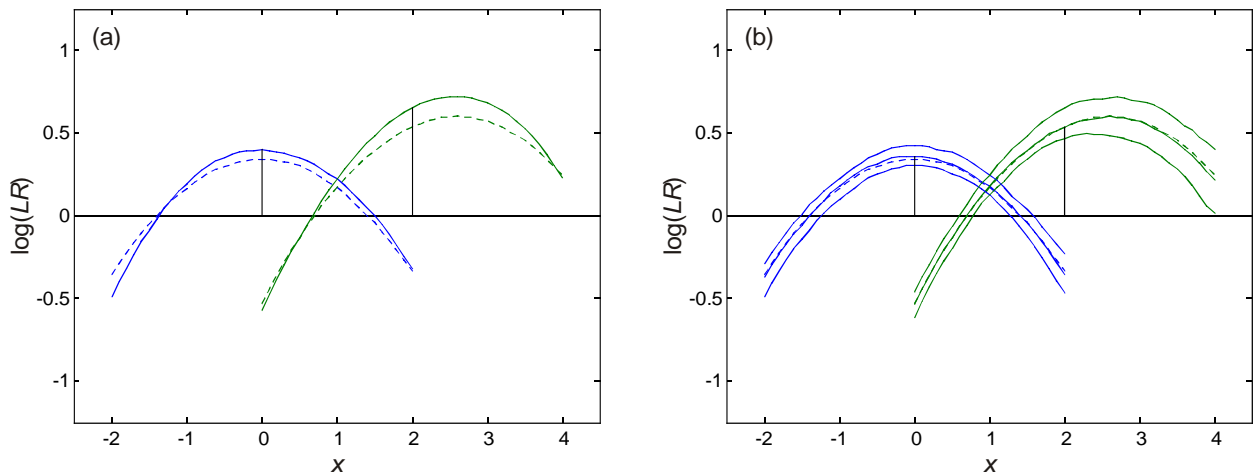

**Figure 2.** Dashed lines: True likelihood-ratio values. (a) Solid lines: Likelihood-ratio values calculated using the pooled-standard-deviation direct method applied to the first set of Monte Carlo sample data. (b) Solid lines: 5th percentile, median, and 95th percentile of likelihood-ratio values calculated using the pooled-standard-deviation direct method applied to 100 sets of Monte Carlo sample data.

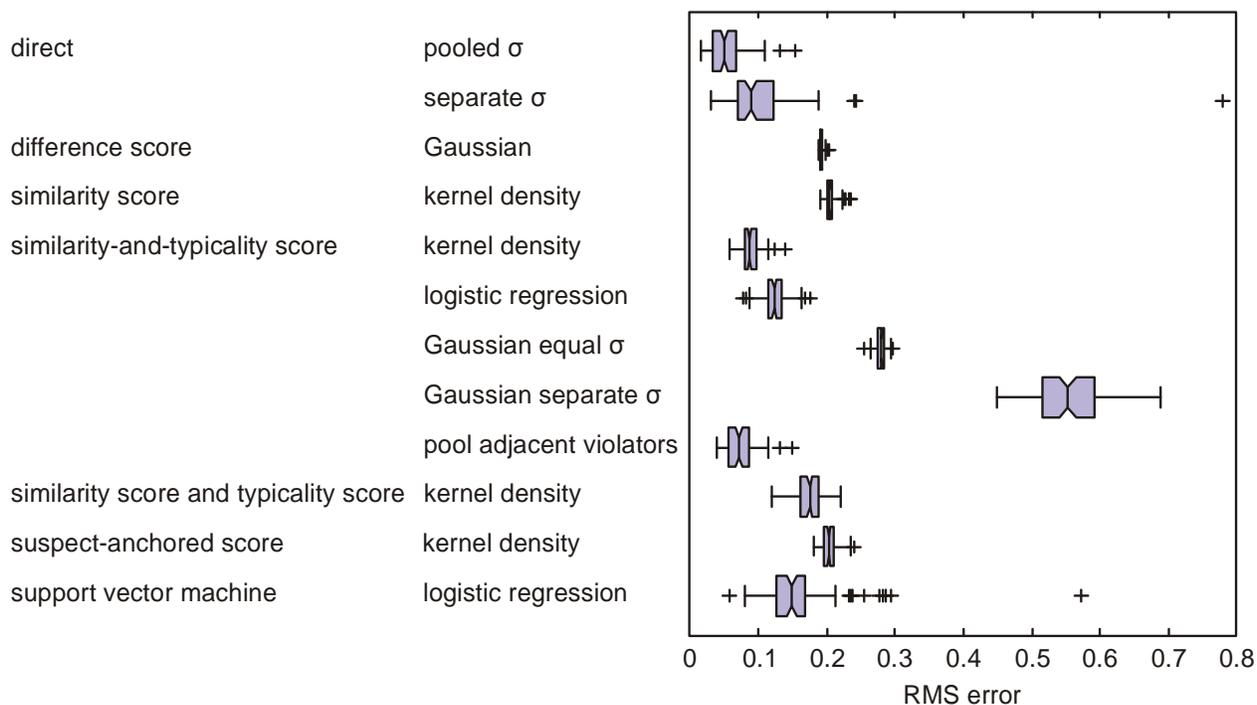

**Figure 3.** RMS error between estimated and true log-likelihood-ratio values calculated over all suspect and offender pairs over 100 sets of samples. Comparisons across direct approaches and approached based on different types of scores (left text column) and different score-to-likelihood-ratio mapping procedures (right text column).

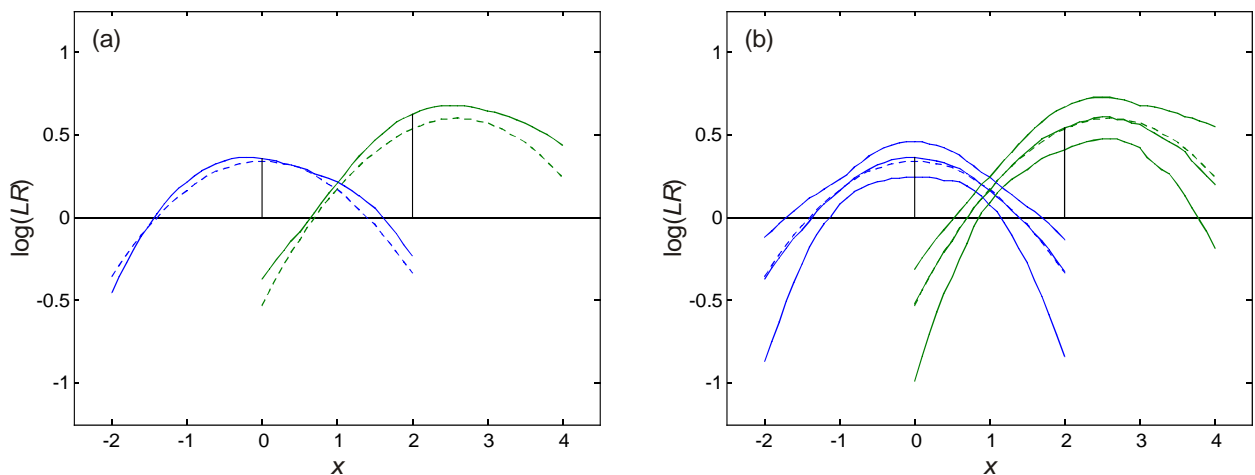

**Figure 4.** Dashed lines: True likelihood-ratio values. (a) Solid lines: Likelihood-ratio values calculated using the separate-standard-deviation direct method applied to the first set of Monte Carlo sample data. (b) Solid lines: 5th percentile, median, and 95th percentile of likelihood-ratio values calculated using the separate-standard-deviation direct method applied to 100 sets of Monte Carlo sample data.

# 5  SCORE-BASED METHODS FOR CALCULATING LIKELIHOOD-RATIO VALUES

Given the simple data structure which could be inferred from the samples employed in the present paper, one would usually simply calculate a likelihood ratio using a direct method and not resort to a scoring method. Scoring methods are more likely to be used when the data are multivariate and potentially multimodal, and the amount of training data is limited with respect to the number of parameters which would have to be calculated if one hoped to produce a relatively low bias low variance model. A scoring method can be considered a means of first projecting a complex multidimensional feature space down into a simpler unidimensional score space, and then calculating a model of the distribution of the univariate scores requiring a much smaller number of parameters to be estimated relative to the amount of training data (see Hepler et al., 2012; Ramos Castro, 2007; Morrison, 2013; Tang & Srihari, 2014). Whereas an attempt to directly model the original feature space may lead to poor parameter estimates and biassed output, a scoring method can ameliorate such biases. The output of the attempted direct model can be treated as scores, and the scores can be converted to likelihood ratios by a second model. One could also consider the output of the attempted direct model to be poorly calibrated likelihood ratios and consider the second model a procedure for calibrating that output, hence the latter procedure is also referred to as "calibration".[7]

The present study focusses on relatively simple distributions, which simplifies exposition. Ultimately the aim would be to address more complex and more realistic problems, but the present study serves as a proof of concept. If a scoring method does not perform well on a simple distribution, then there is little hope that it will perform well on a more complex distribution. In discussion following the presentation of a version of the current paper at the 9th International Conference on Forensic Inference and Statistics, it was suggested that for more complex distributions a similarity-only or difference-only scoring method might actually outperform a similarity-and-typicality method because of difficulty in modelling the typicality part of the latter type of score. Difference-only, similarity-only, and similarity-and-typicality scoring methods applied to a multimodal multivariate Monte Carlo distribution are explored in Appendix A.

Approaches have been proposed which are based on scores taking account of similarities only,

---

[7]Note that I use the term "calibration" to describe a process, whereas Ramos & González-Rodríguez (2013) primarily use it as a noun for what I refer to as a measure of validity (Morrison, 2011).

or based on scores taking account of differences only. As will be illustrated below, in the Monte Carlo simulations such approaches produce likelihood-ratio values which do not pattern after the parallel set of true likelihood-ratio values calculated on the basis of the populations specified for the simulations. As previously discussed in Morrison (2013), the problem with similarity-only scores or difference-only scores is that they do not take account of typicality. Subsequent modelling of the distribution of these scores is unable to introduce an accounting for typicality. Ostensive likelihood ratios calculated on the basis of such models do not, therefore, take account of typicality – they do not take appropriate account of the defence hypothesis. In contrast, scores which take account of both similarity and typicality can theoretically be used to calculate forensically interpretable likelihood ratios. The results of the Monte Carlo simulations below empirically illustrate that an approach based on similarity-and-typicality scores can output likelihood-ratio values which are relatively close to the true likelihood-ratio values for the simulated population.[8]

In discussion following the presentation of a version of the current paper at the 9th International Conference on Forensic Inference and Statistics, it was suggested that for difference-only and similarity-only (and other) scoring methods, the evidence can be defined as the score obtained by comparing the suspect and offender data rather then the measurement(s) made on the offender in the original feature space. A supporting argument was that the likelihood-ratio value will be different if one measures a different property (e.g., length versus width). As mentioned earlier, true likelihood-ratio values, as defined in the present paper, are predicated on knowing the true population distribution for a particular property, and a different value would be expected for a true likelihood-ratio value predicated on knowing the true population distribution for a different property. One must decide what property (or properties) to measure and make it clear that likelihood-ratio values are calculated with respect to that property (or those properties). Using a score rather than original features is not, however, a question of deciding to measure a different property. A score is a derivative of the property which was measured, not a measure of a different property. Scores are derivative and contain less information than was in the original feature space – the original feature values cannot be reconstructed from the derived score values. Difference-only and similarity-only scores are especially lacking in information about typicality in the original feature space. I would argue that the a test of the performance of a procedure for calculating likelihood ratios should compare the output of that procedure with benchmark values calculated using the full information available rather than with benchmark values calculated using reduced information derived form the full information. The full information in the Monte Carlo simulations are the population distributions specified in the feature space. Population distributions in the score space could be derived from these, but they would not contain all the information that was available in the former and thus, I would argue, would not be a suitable benchmark.

---

[8]Morrison (2009) p. 299 described a forensic likelihood ratio as "a *strength-of-evidence* statement in answer to the question: How much more likely are the observed **differences/similarities** between the known and questioned samples to arise under the hypothesis that they have the same origin than under the hypothesis that they have different origins?" (bold added). As I argue in the present paper, I now consider this expression, focussing on similarities or differences, to be incorrect. A more appropriate expression of a forensic likelihood ratio would be something like: "How much more likely are the observed properties of the questioned sample had it come from the same origin as the known sample than had it come from some other member of the relevant population?"

What follows is an illustration of the use of several score-based methods for the calculation of ostensive likelihood ratios. First, examples of systems based on difference-only scores and similarity-only scores are described and their outputs compared with the true likelihood-ratio values for the simulated population. Then, examples of systems based on similarity-and-typicality scores are described and their outputs compared with the true likelihood-ratio values for the simulated population. A similarity score for the numerator and typicality score for the denominator approach, a suspect-anchored approach, and a support-vector-machine approach are also explored. In every case the sample data used to explore the performance of these methods are the same as were used for the direct methods above (some additional data is also used as required by some methods).

## 5.1   Difference-only scores (non-anchored)

Difference scores (aka distance scores, aka dissimilarity scores) quantify the difference between pairs of samples, the larger the score value the larger the difference between the pair. Difference scores could be based on Euclidian distance or city-block distance in the raw feature space, or on Mahalanobis distance taking into account the estimated within-source variance, or on some other transformation of the original feature space. For a recent example of the use of difference-only scores see Vergeer et al. (2014).

Here we illustrate the use of a difference score which is simply the distance between the value of the offender token and the mean of the values of the suspect token in the original feature space. The specified suspect and offender values ($\mu_k$ and $x_q$ respectively) are used for testing the system, and the system is trained on values extracted from the sample data generated in the Monte Carlo simulation. The formula for the difference score is given in Eq. 5:

$$S_{q,k} = g(x_q, \boldsymbol{x}_k) = x_q - \mu_k \qquad\qquad S_{i,j,r} = g(x_{i,r}, \boldsymbol{x}_j) = x_{i,r} - \mu_j \qquad\qquad (5a)$$

$$\mu_j = \frac{1}{T}\sum_{t=1}^{T} x_{j,t} \qquad\qquad (5b)$$

As previously specified, $x_q$ is the value of an offender token and $\mu_k$ is the value of a suspect mean (the latter specified rather than calculated on the basis of an actual sample $\boldsymbol{x}_k$). $x_{i,r}$ is the $r$th token of the $i$th source and $x_{j,t}$ is the $t$th of $T$ tokens of the $j$th source (there are $T = 30$ sample tokens for each source). Note that we keep the sign of the difference from the mean rather than taking the absolute value. This is to simplify training of the score-to-likelihood-ratio model we will use, but ultimately only the magnitude of the score will be important in the mapping function. The signed scores violate the principle that a larger value represent a larger difference, but the unsigned scores (absolute values or magnitudes) conform to this principle.

For training, same-origin scores were calculated by comparing every sample token from a source with the mean of the other 29 tokens from that source ($i = j$, $t \neq r$, $T = 29$). This resulted in a total of 30 scores per source. This was repeated for all 100 sample sources resulting in a total of 3000 same-origin scores. Different-origin scores were calculated as in Eq. 2c, comparing every source with every other source with the constraint $j > i$, for a total of 1950 source pairings [i.e., the upper right of a square matrix excluding the main diagonal, $(100^2 - 100)/2 = 1950$, avoiding the reciprocal lower left], and comparing each token in the first source $i$ with the mean of the tokens $\mu_j$ in the second source $j$ ($T = 30$), for a total of 148 500 different-origin scores.

Neither the suspect data nor the offender datum were involved in the calculation of these training scores. This is referred to as "nonanchord" by Alberink, de Jongh, & Rodríguez (2014), and as "general match" by Hepler et al. (2012) (although in the latter case it was only applied in calculating the denominator). We use the term "non anchored". These scores should be calculated using a database sampled from the relevant population, and so could be considered population anchored. One might a priori assume that because the scores are population anchored, they capture typicality with respect to the relevant population, but as the results below illustrate, they do not adequately reflect typicality for the purposes of calculating forensic likelihood ratios.

The signed difference scores are a linear combination of two normal distributions, and the signed scores therefore have a normal distribution (approximately normal since the simulated population data may vary from perfect normality). The expected value of the mean of both the within-source scores and of the mean of the between-source scores is zero. In practice the calculated means may differ slightly from zero, but we will enforce a mean of zero to guarantee that in score-to-likelihood-ratio modelling a difference score of zero will indicate the least possible difference between an offender token and a suspect mean. This will also guarantee that our model will be symmetrical about zero so that only the magnitude of the difference score is important in mapping from a score to a likelihood ratio. The formulae for mapping from a distance score to a likelihood ratio are given in Eq. 6. Note that this model is designed for the simple Gaussian-distribution data structure arising from our simulated population, and in other scenarios the distributions of the scores may differ.

$$LR_{q,k} = f(S_{q,k} \mid \mathcal{M}_{\mathrm{so}}) \, / \, f(S_{q,k} \mid \mathcal{M}_{\mathrm{do}}), \tag{6a}$$

$$f(S_{q,k} \mid \mathcal{M}_{\mathrm{so}}) = \varphi(\, S_{q,k} \mid 0, \, \sigma_{\mathrm{so}} \,), \tag{6b}$$

$$f(S_{q,k} \mid \mathcal{M}_{\mathrm{do}}) = \varphi(\, S_{q,k} \mid 0, \, \sigma_{\mathrm{do}} \,), \tag{6c}$$

$$\sigma_{so} = \sqrt{\frac{1}{U-1} \sum_{u=1}^{U} S_{so,u}^{2}} \;\;, \tag{6d}$$

$$\sigma_{do} = \sqrt{\frac{1}{V-1} \sum_{v=1}^{V} S_{\mathrm{do},v}^{2}} \;\;, \tag{6e}$$

where $S_{so}$ and $S_{do}$ are same-origin and different-origin training scores respectively, and $U$ and $V$ are the number of same-origin and different-origin training scores respectively. $\varphi$ is a normal distribution, in this case with a mean of zero. $\sigma_{so}$ and $\sigma_{do}$ are the standard deviations of the same-origin and different-origin training scores respectively, each centred around a mean of zero.

The modelled same-origin and different-origin distributions are shown in Fig. 5a (based on scores derived from the first sample set drawn from the simulated population). Fig. 5b shows the resulting score to log likelihood ratio mapping function, which is directly derived from the relative heights of the curves shown in Fig. 5a. The sign of the score is irrelevant for score-to-likelihood-ratio mapping, so Fig. 5b shows absolute score values on the abscissa.

Fig. 6 shows plots of likelihood-ratio values calculated for the specified suspect means and offender token values ($\mu_k$ and $x_q$) using the method based on difference-only scores and Gaussian models of score distributions. Note that because difference-only scores do not take account of the typicality of the offender tokens with respect to the relevant population, the likelihood-ratio values related to the rightmost suspect in Fig. 6 do not show the same asymmetry as occurs for the true likelihood-ratio values.[9] These likelihood-ratio values calculated using the difference-only scores are not close to the true likelihood-ratio values: the median RMS error value was 0.192, see Fig. 3 for a visual comparison of RMS error values across different approaches. This empirically illustrates that modelling the distribution of same-origin and different-origin difference-only scores is not an appropriate method for calculating forensic likelihood-ratio values.[10]

---

[9]The pattern of these results is very similar to those for the comparison of score-based calculations with true likelihood-ratio values in Hepler et al. (2012): Compare the top right panel of Fig. A1 in Hepler et al. with the leftmost set of suspect and offender values in Fig. 6a in the present paper, and the top middle panel of Fig. A1 in Hepler et al. with the rightmost set of suspect and offender values in Fig. 6a in the present paper. None of the three score-based methods tested in Hepler et al. appear to have adequately captured typicality. All three methods used suspect-anchored scores for the numerator. Scores for the denominator were either non anchored, offender anchored, or suspect anchored. Scores were squared differences and the scores were modelled using chi-squared distributions.

[10]Tang & Srihari (2014) have recently proposed a procedure for combining distance-only likelihood ratios (or similarity-only likelihood ratios) with a factor which takes account of typicality, but this particular approach is not explored in the present paper.

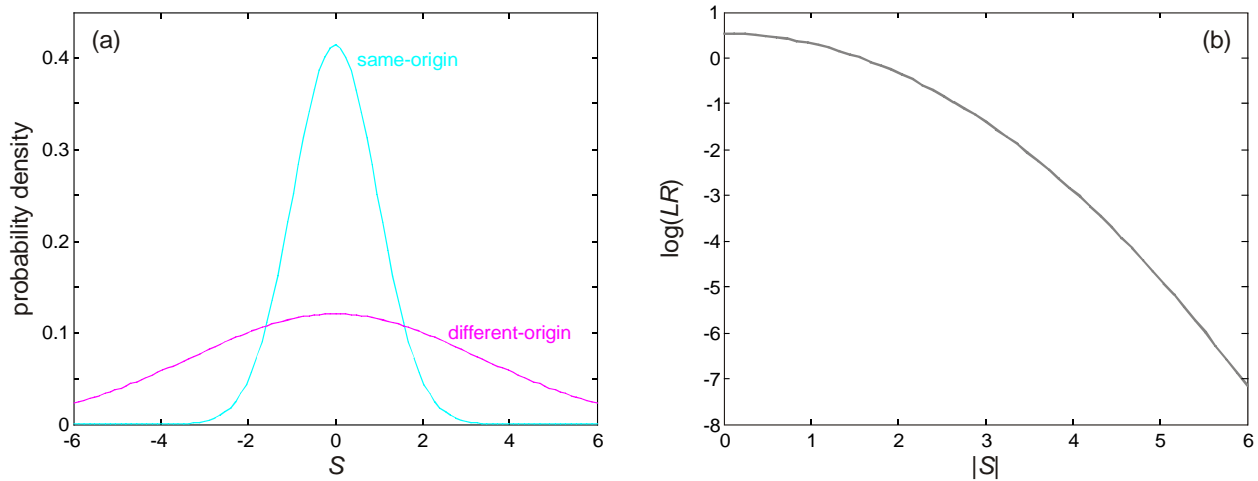

**Figure 5.** (a) Modelled distributions of same-origin and different-origin difference scores (Gaussian models). (b) Absolute score value to $\log_{10}$ likelihood-ratio value mapping based on the distributions shown in Fig. 5a.

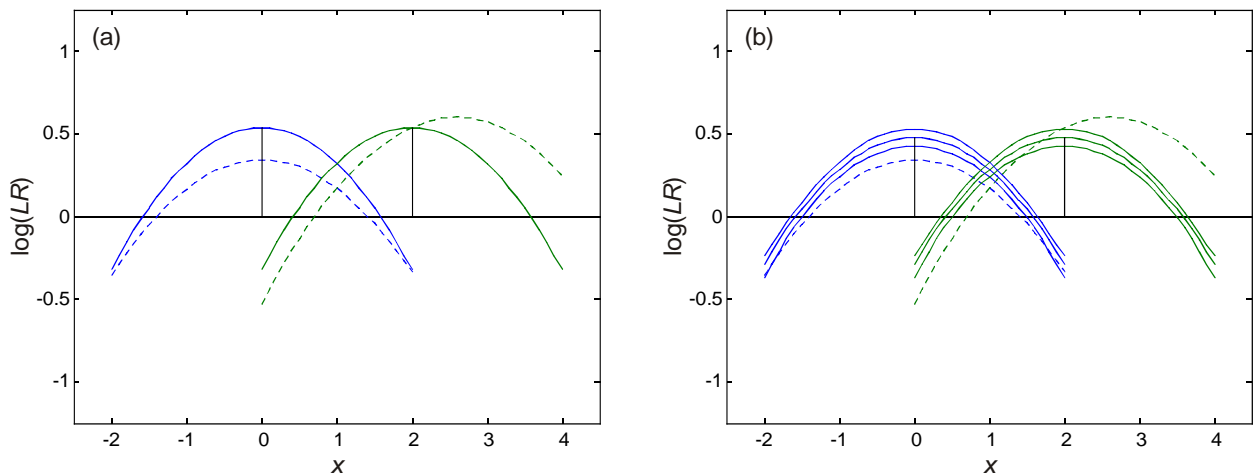

**Figure 6.** Dashed lines: True likelihood-ratio values. (a) Solid lines: Likelihood-ratio values calculated using difference scores plus Gaussian modelling applied to the first set of Monte Carlo sample data. (b) Solid lines: 5th percentile, median, and 95th percentile of likelihood-ratio values calculated using difference scores plus Gaussian modelling applied to 100 sets of Monte Carlo sample data.

## 5.2 Similarity-only scores (non-anchored)

Similarity scores are in a sense the inverse of difference scores, they quantify the similarity between pairs of samples, the larger the score value the smaller the difference between the pair. A similarity score could literally be the inverse of a difference score, or could, for example, be based on the likelihood of a point value given a probability-density-function model, or on the correlation between two sets of samples in a multivariate feature space. For a recent example of the use of similarity-only scores see van Houten et al. (2011). Here we illustrate the use of similarity-only scores based on likelihoods from normal-distribution models.

As with the earlier models, the specified suspect and offender values ($\mu_k$ and $x_q$ respectively) are used for testing the system, and the system is trained on values extracted from the sample data generated in the Monte Carlo simulation. The formula for the similarity score is given in Eq. 7:

$$S_{q,k} = g(x_q, \boldsymbol{x}_k) = \varphi(\, x_q \mid \mu_k, \, \sigma_k\,), \tag{7a}$$

$$S_{i,j,r} = g(x_{i,r}, \boldsymbol{x}_j) = \varphi(\, x_{i,r} \mid \mu_j, \, \sigma_j\,), \tag{7b}$$

where $\varphi$ is a normal distribution, $x_q$, $\mu_k$, $x_{i,r}$, and $\mu_j$ are as previously defined, and $\sigma_k$ and $\sigma_j$ are the estimate of the pooled within-source standard deviation, $\sigma_k = \sigma_j = \sigma$ for both $k$ and all $j$. 3000 same-origin training scores and 148 500 different-origin training scores were calculated in the same manner as for the difference-only scores above.

The resulting scores have distributions for which there is no obvious parametric model. We therefore model them using kernel-density models. The probability density is calculated via Eq. 8 (see Bowman & Azzalini, 1997, p. 3; Hastie, Tibshirani, & Friedman, 2009, p. 209):

$$f(S) = \frac{1}{A} \sum_{a=1}^{A} \varphi\big(S - S_a \,|\, 0, \sigma_b\big), \tag{8}$$

where $S$ is the score value at which the density is to be evaluated, and $S_a$ is one of $A$ training scores calculated as in Eq. 7b. A Gaussian kernel was used with a mean of 0 and a standard deviation (or bandwidth) of $\sigma_b = 0.02$. The bandwidth was chosen somewhat arbitrarily on the basis of eyeballing probability-density plots and selecting a value which seemed to give a reasonable degree of smoothing, i.e., not too noisy but preserving major aspects of the distributions' shapes. Fig. 7a shows plots of the modelled distributions of same-origin scores and different-origin scores, and Fig. 7b shows the resulting score to log likelihood ratio mapping.

Fig. 8 shows plots of likelihood-ratio values calculated using the method based on similarity-only scores and kernel-density models of score distributions. Choosing a different degree of smoothing for the kernel-density models would have produced different results, but we delay discussion of potential problems related to kernel-density models until later. At present we note that, as with the distance-only scores, because similarity-only scores do not take account of the typicality of the offender tokens with respect to the relevant population, the likelihood-ratio values related to the rightmost suspect in Fig. 8 do not show the same asymmetry as occurs for the true likelihood-ratio values. These likelihood-ratio values calculated using similarity-only scores are not close to the true likelihood-ratio values: the median RMS error value was 0.204, see Fig. 3 for a visual comparison of RMS error values across different approaches. This empirically illustrates that modelling the distribution of same-origin and different-origin similarity-only scores is not an appropriate method for calculating forensic likelihood-ratio values.

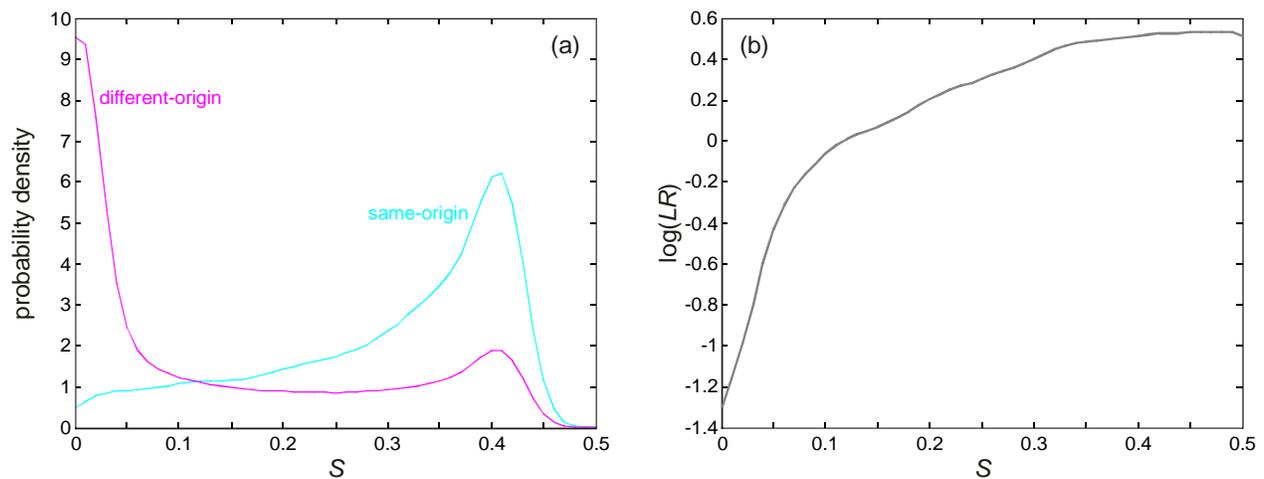

**Figure 7.** (a) Modelled distributions of same-origin and different-origin similarity scores (kernel-density models). (b) Score-to-$\log_{10}$-likelihood-ratio mapping based on the distributions shown in Fig. 7a.

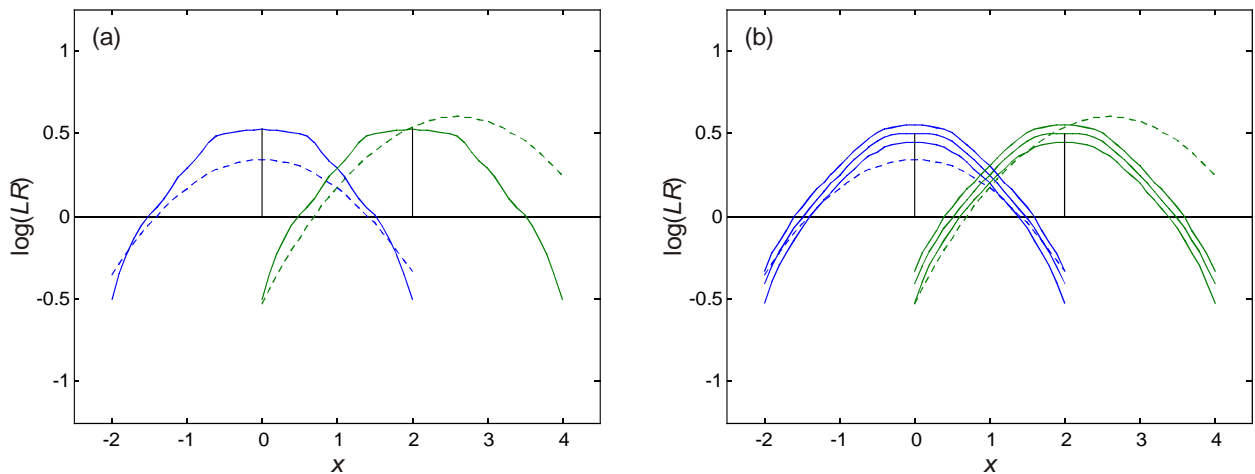

**Figure 8.** Dashed lines: True likelihood-ratio values. (a) Solid lines: Likelihood-ratio values calculated using similarity scores plus kernel density modelling applied to the first set of Monte Carlo sample data. (b) Solid lines: 5th percentile, median, and 95th percentile of likelihood-ratio values calculated using similarity scores plus kernel density modelling applied to 100 sets of Monte Carlo sample data.

### 5.3 Similarity-and-typicality scores (non-anchored)

An attempt to implement a direct method for the calculation of likelihood ratios may result in poor output if the amount of training data is small compared to the number of parameters for which values need to be estimated. In such a case the output can be considered to be poorly calibrated likelihood-ratio values which should be calibrated before being interpreted, or can be considered scores which should be converted to likelihood ratios before being interpreted. The likelihood-ratio values calculated using the pooled-standard-deviation direct method above were intrinsically well calibrated, but in order to illustrate calibration we will treat their logarithms as similarity-and-typicality scores which we will then convert to likelihood-ratio values. The procedures for converting from scores to likelihood ratios illustrated in this section are kernel density models, logistic-regression calibration, Gaussian models with equal variances, Gaussian models with separate variances, and the non-parametric pool-adjacent-violations algorithm. All of these were previously illustrated in Ramos Castro (2007) and the latter four in Brümmer, Swart, & van Leeuwen (2014) but not with comparison to true likelihood-ratio values, and in the latter case not in the context of forensic application.

Similarity-and-typicality scores must take into account the similarity of the offender sample with respect to the suspect sample while at the same time taking into account the typicality of the offender sample with respect to a sample of the relevant population. The similarity-and-typicality scores we use in the present paper are calculated as in Eq. 9:

$$S_{q,k} = g\left(x_q \mid \mu_k, \sigma_k, \boldsymbol{B}\right) = \log\left(\varphi\left(x_q \mid \mu_k, \sigma_k\right)\right) - \log\left(\frac{1}{H}\sum_{h=1}^{H}\varphi\left(x_q \mid \mu_h, \sigma_h\right)\right), \tag{9a}$$

$$S_{i,j,r} = g\left(x_{i,r} \mid \mu_j, \sigma_j, \boldsymbol{B}\right) = \log\left(\varphi\left(x_{i,r} \mid \mu_j, \sigma_j\right)\right) - \log\left(\frac{1}{H}\sum_{h=1}^{H}\varphi\left(x_{i,r} \mid \mu_h, \sigma_h\right)\right), \tag{9b}$$

$$\mu_h = \frac{1}{T}\sum_{t=1}^{T} x_{h,t} \ , \tag{9c}$$

where $\boldsymbol{B}$ is the background sample. For the calculation of a score for a suspect and offender pair $S_{q,k}$, $\boldsymbol{B}$ consisted of data from all 100 sampled sources ($H = 100$). For the calculation of training scores $S_{i,j,r}$, $\boldsymbol{B}$ consisted of data from all 100 sampled sources except for sources $i$ and $j$ which were being compared (i.e., a cross-validation procedure was adopted, and if $i = j$, $H = 99$, else if $i \neq j$, $H = 98$). We use the pooled within-source standard deviation throughout, $\sigma_k = \sigma_j = \sigma_h = \sigma$ for both $k$, all $j$, and all $h$. Note that the calculation of scores in Eq. 9 parallels the calculation of likelihood-ratio values in the direct method given in Eq. 3 except that the calculations are performed and the output saved in a logarithmic space. Any base can be used for the logarithm. For convenience all logarithmic results in the present paper are presented using base-ten logarithms.

3000 same-origin training scores and 148 500 different-origin training scores were calculated in the same manner as for the distance-only scores above (for same origin $i = j$, $t \neq r$, $T = 29$; and for different origin $j > i$, $T = 30$).

As discussed in Morrison (2013), similarity-and-typicality scores are similar to likelihood ratios in that they take account of both the similarity of the offender with respect to the suspect and the typicality of the offender with respect to the relevant population, but their absolute values are not interpretable. Given a lower and a higher valued score, they could both correspond to likelihood ratios which are less than one, or both correspond to likelihood ratios which are greater than one, or the lower of the two to less than one and the higher of the two to greater than one. Scores which are close in value could lead to likelihood ratios which are far apart in value, or scores which are far apart in value could lead to likelihood ratios which are close in value. Only the rank order of similarity-and-typicality scores have meaning: a higher valued score must correspond to a higher valued likelihood ratio than a lower valued score. A function for mapping between a score and a likelihood ratio should therefore preserve the rank ordering of scores – the function should be monotonic. In addition, in order for the output of the mapping function to be interpretable as a likelihood ratio it should be the result of either a model which directly evaluates the ratio of two likelihoods or the result of a model that is clearly analogous to a model which directly evaluates the ratio of two likelihoods.

### 5.3.1 Kernel-density models

Fig. 9a shows plots of kernel distributions fitted to the same-origin training scores and to the different-origin training scores (scores from the first sample set drawn from the simulated population). The probability densities were calculated using Eq. 8. The bandwidth, $\sigma_b = 0.2$, was chosen somewhat arbitrarily on the basis of eyeballing probability-density plots and selecting a value which seemed to give a reasonable degree of smoothing. The distribution of same-origin scores is relatively symmetric, whereas that for different-origin scores is heavily skewed towards lower values (the abscissa in Fig. 9a is truncated at −3).

Fig. 10 shows the results of calculating likelihood-ratio values for the suspect and offender pairs on the basis of these kernel-density models. The median RMS error value was 0.088, see Fig. 3 for a visual comparison of RMS error values across different approaches. Note that this approach using similarity-and-typicality scores and converting the scores to likelihood ratios using kernel density models produces results which are close to the true likelihood-ratio values – much closer than the approaches using distance-only or similarity-only scores, and only a little worse than the pooled-standard-deviation direct approach. Empirically, this approach therefore appears to be an appropriate method for calculating forensic likelihood-ratio values.

A potential problem with the use of kernel-density models to convert from scores to likelihood ratios is that an appropriate degree of smoothing has to be selected. Too little smoothing could lead to erratic fluctuations in the likelihood-ratio value given small changes in score value (see Fig. 11a based on a bandwidth of 0.02), and too much smoothing could lead to models for which the likelihood-ratio output is close to 1 (log likelihood ratio of 0) across a broad range of score values (see Fig. 11b based on a bandwidth of 2). Fig. 11 is based on rather extreme values to clearly illustrate the point. Another problem is that a higher-valued score should always result in a higher-valued likelihood ratio than a lower-valued score, but such a monotonic relationship is not guaranteed by the use of kernel-density models. When using a kernel-density approach one would have to be careful to make sure that monotonicity holds across any score values that the models are ever expected to be applied to. Fig. 9b (solid line) shows the score to log-likelihood-ratio mapping function for the kernel-density models with a bandwidth of 0.2. Log-likelihood-ratio values rise in response to increases in score values, but then drop precipitously a little above a score value of 2 (the highest value of a same-origin training score was 2.15). This is not necessarily a problem if the model will only ever be exposed to scores within the range for which monotonicity holds (as is the case for our test data here), but one would have to be careful to make sure this is the case. The potential for violations of monotonicity will increase if the amount of training data is reduced or the degree of smoothing is reduced. In general, kernel density models perform poorly if they are applied to areas which are not sampled or which are sparsely sampled in the training data.

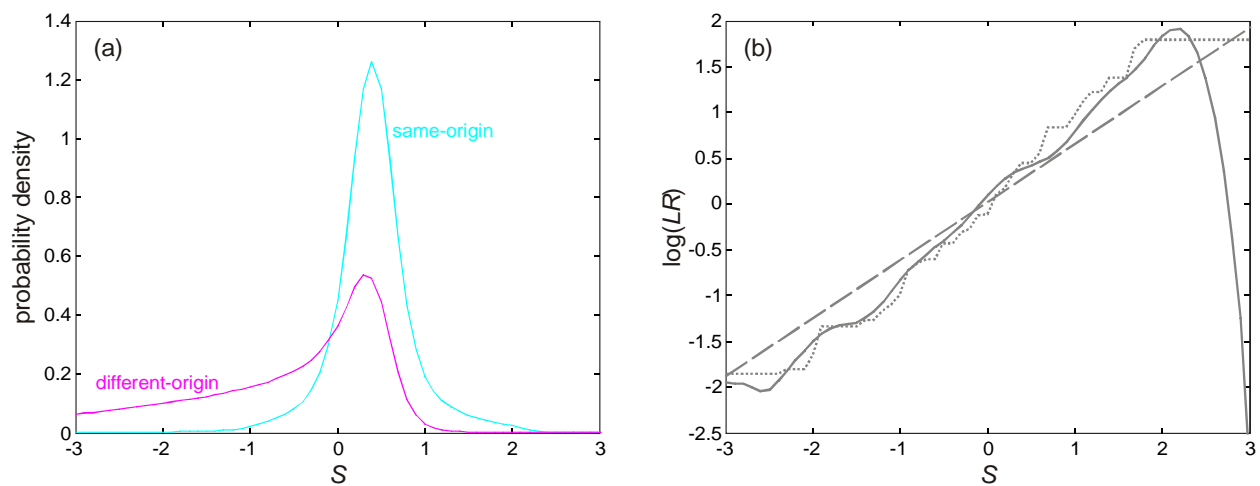

**Figure 9.** (a) Modelled distributions of same-origin and different-origin similarity-and-typicality scores (kernel-density models). (b) Score-to-log$_{10}$-likelihood-ratio mapping based on the distributions shown in Fig. 9a. Solid line: kernel-density mapping. Dashed line: logistic-regression mapping. Dotted line: Pool-adjacent-violators mapping

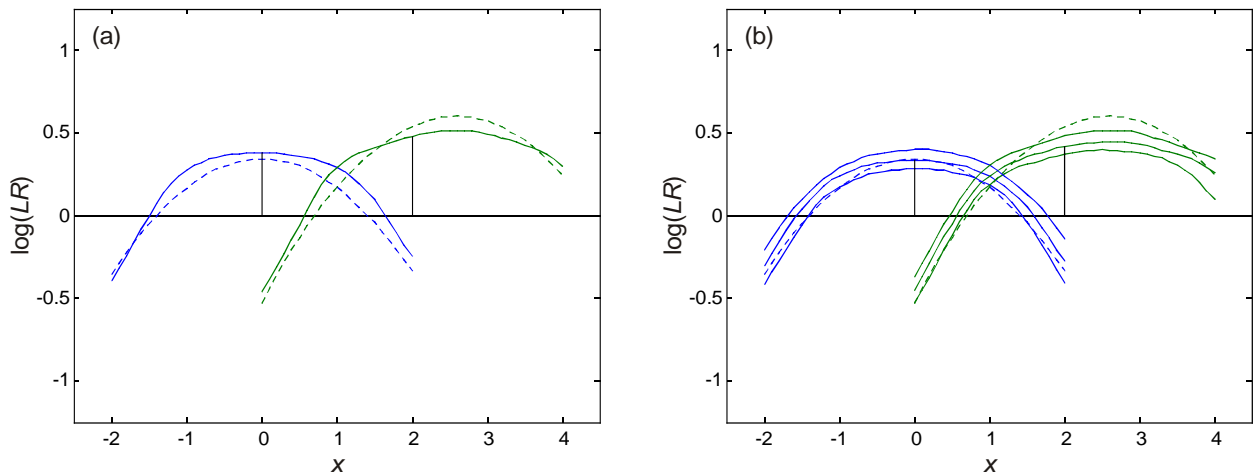

**Figure 10.** Dashed lines: True likelihood-ratio values. (a) Solid lines: Likelihood-ratio values calculated using similarity-and-typicality scores plus kernel density modelling applied to the first set of Monte Carlo sample data (kernel-density bandwidth of 0.2). (b) Solid lines: 5th percentile, median, and 95th percentile of likelihood-ratio values calculated using similarity-and-typicality scores plus kernel density modelling applied to 100 sets of Monte Carlo sample data.

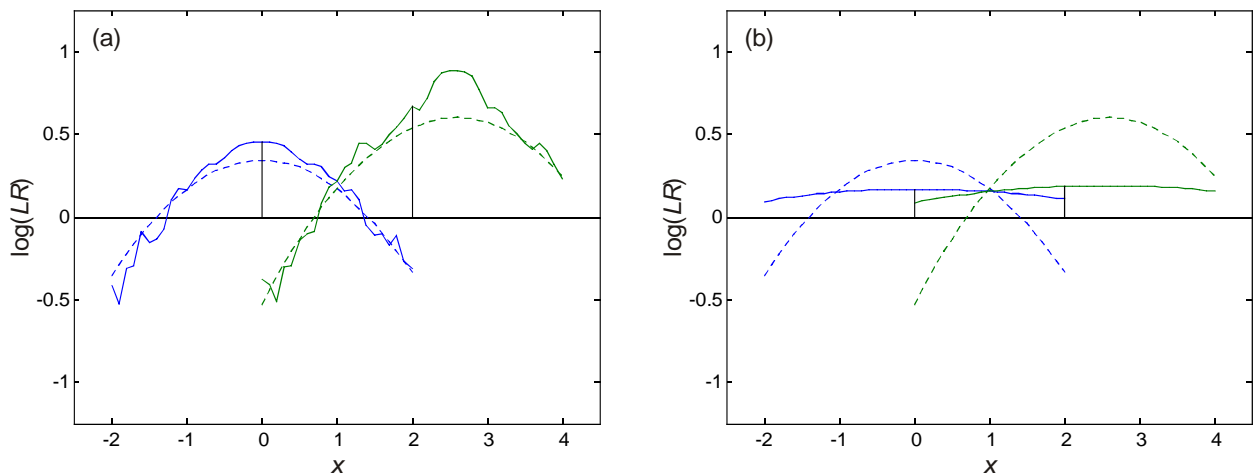

**Figure 11.** Dashed lines: True likelihood-ratio values. Solid lines: Likelihood-ratio values calculated using similarity-and-typicality scores plus kernel density modelling applied to the first set of Monte Carlo sample data. (a) Kernel density bandwidth of 0.02. (b) Kernel density bandwidth of 2.

*5.3.2 Logistic-regression calibration*

A standard approach for converting similarity-and-typicality scores to likelihood ratios is logistic-regression calibration (Brümmer & du Preez, 2005; González-Rodríguez et al., 2007; Ramos Castro, 2007; Morrison, 2013; van Leeuwen & Brümmer, 2013; Mandasari et al., 2014; Ali, 2014). Tautologically, in order for a likelihood ratio to be a likelihood ratio it should be the ratio of two likelihoods. All the approaches for converting scores to likelihood ratios described so far fit a generative model to the same-origin training scores and a generative model to the different-origin training scores, then take the ratio of the probability density of the models evaluated at the value of the score from the comparison of the suspect and offender samples, i.e., the ratio of the likelihood of the same-origin model evaluated at the suspect-and-offender score and the likelihood of the different-origin model evaluated at the suspect-and-offender score. In contrast, logistic regression is a discriminative procedure rather than a generative procedure. It does not directly evaluate the ratio of two likelihoods. As discussed in Morrison (2013), however, logistic regression is analogous to a generative model: A logistic regression model with equal priors for each category (same-origin category and different-origin category in this case) gives the same results as a model in which the same-origin scores are modelled by a single Gaussian and the different-origin scores are modelled by a single Gaussian and the variances of the two Gaussians are equal, if the assumptions of normality and equal variance hold. Because it is analogous to a generative model and the output of the generative model is clearly the ratio of two likelihoods, the output of a logistic-regression model can be interpreted as if it were the ratio of two likelihoods – it can be interpreted as a likelihood ratio. An advantage of using a logistic-regression model or an equal-variance Gaussian model over a kernel-density model or a separate-variance Gaussian model is that the former two, unlike the latter two, guarantee monotonicity; in fact they result in a linear mapping between score values and log-likelihood-ratio values (see Morrison, 2013; van Leeuwen & Brümmer, 2013). An advantage of using a logistic-regression model rather than an equal-variance Gaussian model is that the former is more robust to violations of the assumptions of normality and equal variance (as illustrated below).

The logistic-regression model for converting a score to a log-likelihood ratio is simply a linear transformation as given in Eq. 10, in which $\alpha$ is an intercept parameter and $\beta$ a slope parameter. Values for these parameters are trained via an iterative procedure in a logged-odds space (the similarity-and-typicality scores from Eq. 9 have the form of logged odds and log likelihood ratios have the form of logged odds). The procedure is outlined in various textbooks (e.g., Pampel, 2000, p. 44–45; Hosmer & Lemeshow, 2000, p.7–10) and in greater algorithmic detail in various other textbooks (e.g., Hastie, Tibshirani, & Friedman, 2009, p. 120–122), and we do not discuss the details here.

$$\log(LR) = \alpha + \beta S \qquad (10)$$

A logistic-regression model was trained, coding the same-origin training scores with a category value of 1 and different-origin training scores with a category value of 0, and giving equal weight to each category (the calculations were performed using the `train_llr_fusion` function in the FOCAL TOOLKIT; Brümmer, 2005). The $\alpha$ and $\beta$ coefficient values from this model were then used to transform the suspect-and-offender scores to likelihood ratios. The score-to-log-likelihood-ratio mapping function for the first sample set drawn from the simulated population is shown in Fig 9b (dashed line). The likelihood-ratio results are shown in Fig. 12. The median RMS error value was 0.124, see Fig. 3 for a visual comparison of RMS error values across different approaches. Note that this approach using similarity-and-typicality scores and converting the scores to likelihood ratios using logistic regression produces results which may be considered reasonably close to the true likelihood ratios – they are not as close as the results from the kernel density model, but the logistic-regression approach has the advantages that it does not require the selection of a bandwidth parameter and monotonicity is guaranteed. Theoretically and empirically, this approach therefore appears to be an appropriate method for calculating forensic likelihood-ratio values.

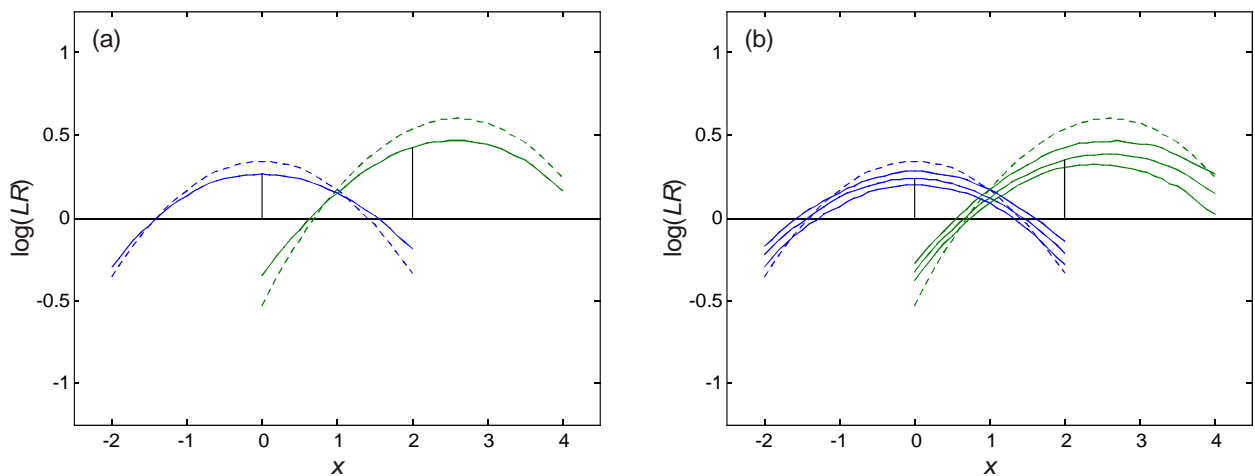

**Figure 12.** Dashed lines: True likelihood-ratio values. (a) Solid lines: Likelihood-ratio values calculated using

similarity-and-typicality scores plus logistic-regression calibration applied to the first set of Monte Carlo sample data.

(b) Solid lines: 5th percentile, median, and 95th percentile of likelihood-ratio values calculated using similarity-and-typicality scores plus logistic-regression calibration applied to 100 sets of Monte Carlo sample data.

### 5.3.3 Gaussian approach – equal variances and separate variances

The robustness of the logistic-regression approach is quite remarkable given that the distributions of the scores shown in Fig. 9a are clearly not normal, do not have equal variance, and the distribution of different-origin scores is extremely skewed. There is some effect: calculated likelihood-ratio values are on average closer to 1 (log-likelihood-ratio values closer to 0) than the true likelihood-ratio values.

Although logistic regression is justified as a procedure for converting scores to likelihood ratios because is analogous to an equal-variance Gaussian model, the latter model is not robust to the extreme violations of normality and equal variance observed for the similarity-and-typicality scores in this case. To illustrate this, the suspect-and-offender scores were converted to likelihood ratios using two Gaussians, one with its mean calculated on the basis of the 3000 same-origin training scores and the other with its mean calculated on the basis of the 148 500 different-origin training scores and both with the same pooled standard deviations calculated using all training scores. The formulae for the calculations are shown in Eq. 11, where $\mu_{so}$ and $\mu_{do}$ are the means of the same-origin and different-origin training scores respectively, and $\sigma_{so} = \sigma_{do} = \sigma$ is the pooled standard deviation calculated from both groups of training scores with equal weight given to each category rather than weighting according to the number of tokens in each category ($U$ = number of same-origin scores and $V$ = number of different origin scores). Like logistic regression, the score-to-log-likelihood-ratio mapping for the equal-variance Gaussian approach reduces to a linear function (see Morrison, 2013; van Leeuwen & Brümmer, 2013).

$$LR = f(S_{q,k} \mid H_{so}) \, / \, f(S_{q,k} \mid H_{do}) \tag{11a}$$

$$f(S_{q,k} \mid H_{so}) = \varphi(S_{q,k} \mid \mu_{so}, \sigma_{so}) \tag{11b}$$

$$f(S_{q,k} \mid H_{do}) = \varphi(S_{q,k} \mid \mu_{do}, \sigma_{do}) \tag{11c}$$

$$\varphi\left(S\,|\,\mu,\sigma\right) = \frac{1}{\sigma\sqrt{2\pi}}\,e^{\frac{-(S-\mu)^2}{2\sigma^2}} \tag{11d}$$

$$\sigma = \sqrt{\frac{\displaystyle\sum_{u=1}^{U}\left(S_{\mathrm{so},u}-\mu_{\mathrm{so}}\right)^2 + \sum_{v=1}^{V}\left(S_{\mathrm{do},v}-\mu_{\mathrm{do}}\right)^2}{U+V-2}} \tag{11e}$$

The results of converting the similarity-and-typicality scores to likelihood ratios using the equal-variance Gaussian approach are shown in Fig. 13. The median RMS error value was 0.281, see Fig. 3 for a visual comparison of RMS error values across different approaches. The likelihood ratios calculated using the equal-variance Gaussian model deviate substantially from the true likelihood ratios for these Monte Carlo simulations. Because of the large spread in the different-origin scores, the calculated pooled standard deviation is large and this results in all the calculated likelihood ratios being close to a likelihood-ratio value of 1 (log likelihood ratio of 0). The equal-variance Gaussian approach is therefore not recommended as a procedure for converting scores to likelihood ratios.

van Leeuwen & Brümmer (2013) count as an advantage the fact that the equal-variance Gaussian approach has a closed-form solution which can be trained faster than the iterative procedure used for logistic regression, but they also note that it will not work well if the distributions of the scores violate the assumption of normality. Nor do we expect it to work well if the assumption of equal variance is violated.

One might ask whether, given the apparent difference in the variances of the same-origin scores and the different-origin scores, a model using separate variances for each Gaussian might ameliorate the problem seen with the application of the equal-variance model in this case. Fig. 14 shows the results of using a separate-variance Gaussian model, identical to the equal-variance Gaussian model except that in Eq. 11 $\sigma_{\mathrm{so}}$ was calculated on the basis of same-origin scores only and $\sigma_{\mathrm{do}}$ was calculated on the basis of different-origin scores only. The median RMS error value was 0.552, see Fig. 3 for a visual comparison of RMS error values across different approaches. The calculated likelihood ratios are generally very far from the true likelihood ratios. The score-to-likelihood-ratio mapping function is not monotonic: Log likelihood ratios initially increase as scores increase, but beyond a score value of approximately 0.4 (just after the mean of the same-origin scores) the same-origin distribution function drops faster than the different-origin distribution function, and log likelihood ratios decrease as scores increase (see Fig. 15). Botti, Alexander, & Drygajlo (2004) and van Houten et al. (2011) also encountered this problem and those papers include figures showing score to log-likelihood-ratio mappings similar to Fig. 15 of the present paper.

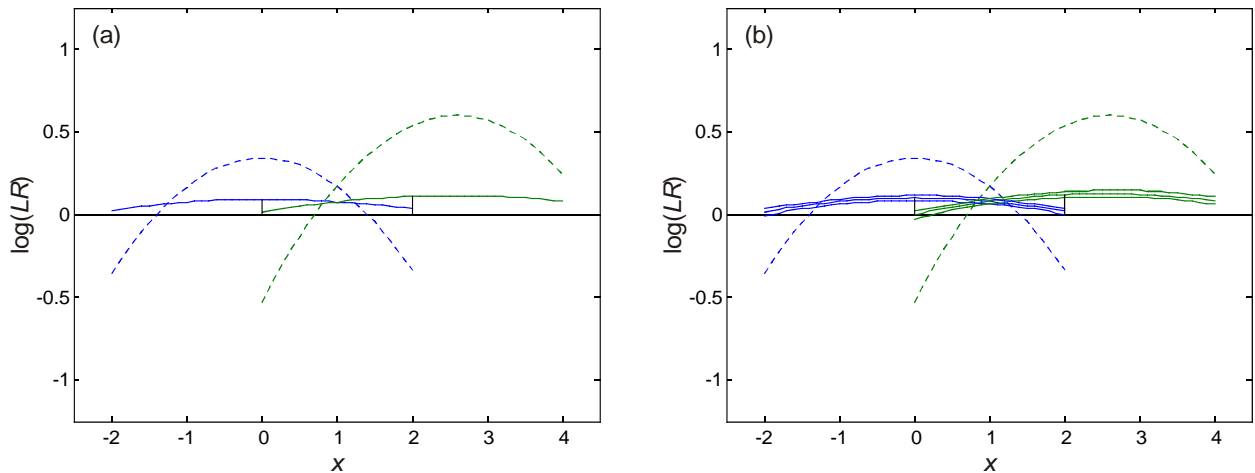

**Figure 13.** Dashed lines: True likelihood-ratio values. (a) Solid lines: Likelihood-ratio values calculated using similarity-and-typicality scores plus equal-variance Gaussian modelling applied to the first set of Monte Carlo sample data. (b) Solid lines: 5th percentile, median, and 95th percentile of likelihood-ratio values calculated using similarity-and-typicality scores plus equal-variance Gaussian modelling applied to 100 sets of Monte Carlo sample data.

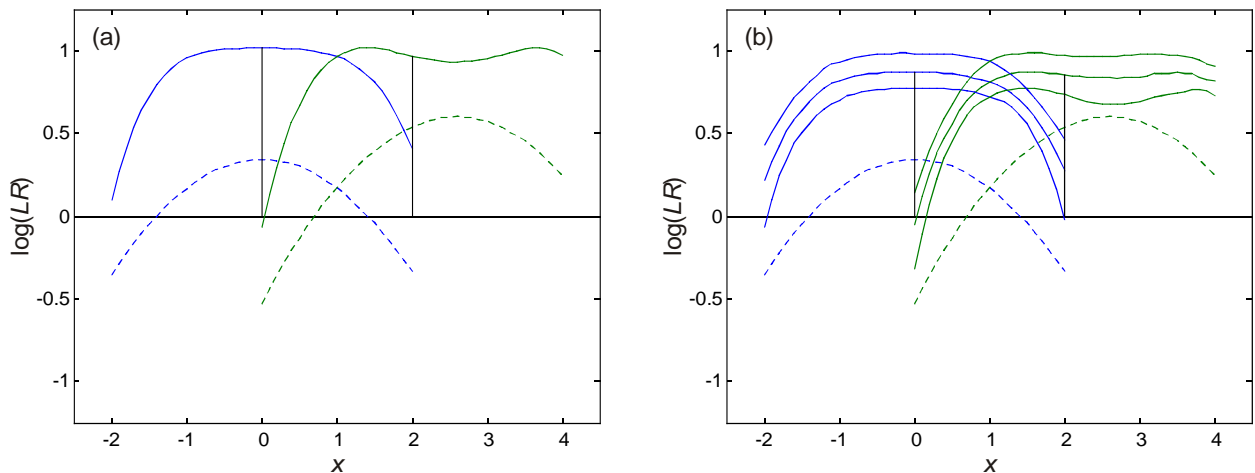

**Figure 14.** Dashed lines: True likelihood-ratio values. (a) Solid lines: Likelihood-ratio values calculated using similarity-and-typicality scores plus separate-variance Gaussian modelling applied to the first set of Monte Carlo sample data. (b) Solid lines: 5th percentile, median, and 95th percentile of likelihood-ratio values calculated using similarity-and-typicality scores plus separate-variance Gaussian modelling applied to 100 sets of Monte Carlo sample data.

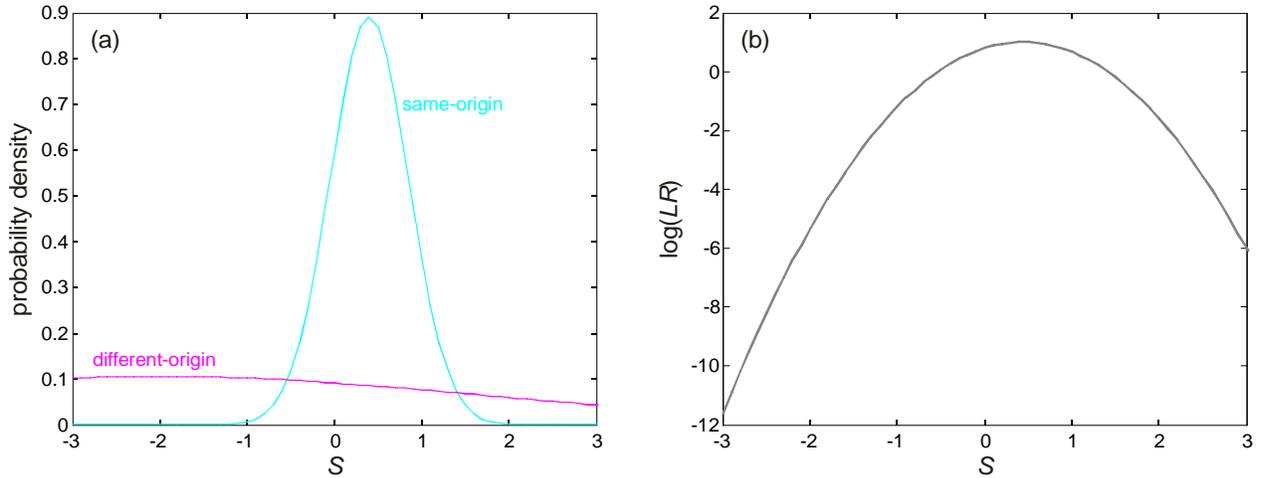

**Figure 15.** (a) Modelled distributions of same-origin and different-origin similarity-and-typicality scores (separate-variance Gaussian models). (b) Score-to-log$_{10}$-likelihood-ratio mapping based on the distributions shown in Fig. 15a.

### 5.3.4 Pool adjacent violators

Another procedure which could be used for converting scores to likelihood ratios is the non-parametric pool-adjacent-violators algorithm (Zadrozny & Elkan, 2002; Brümmer & du Preez, 2006; van Leuwen & Brümmer, 2007; Ramos & González-Rodríguez, 2013; Ali et al., 2013; Ali, 2014). The procedure is as follows:

1. All training scores are ranked in ascending order.

2. Each position in the rank is assigned a value of 0 or 1. Rank positions corresponding to different-origin scores are assigned a value of 0 and those corresponding to same-origin scores are assigned a value of 1.

3. Groups of adjacent rank positions which violate monotonicity, e.g., $S_{a-1} > S_a$, are pooled together and all members of the group are given the same new value which is the mean of the old values of the members of the group, e.g., $S'_{a-1} = S'_a = (S_{a-1} + S_a)/2$.

4. The algorithm is first applied to adjacent pairs of rank positions then iteratively applied to include groups of already pooled rank positions, e.g.:

|   |      | 0 | 0 | 1    | 1    | 0    | 1 | 1 |
|---|------|---|---|------|------|------|---|---|
| → |      | 0 | 0 | 1    | 0.5  | 0.5  | 1 | 1 |
| → |      | 0 | 0 | 0.67 | 0.67 | 0.67 | 1 | 1 |

5. The resulting values are treated as posterior probabilities and converted to log-odds: $\log(p) - \log(1-p)$

6. The prior odds given the relative number of same- and different-origin scores, $\log(U/V)$, is subtracted to arrive at log likelihood ratios.

If there is a mismatch between the number of different-origin scores and the number of same-origin scores used in training, the posterior odds will be biassed by this mismatch. Step 6 corrects the bias. The outcome is equivalent to giving equal priors to the same-origin and different-origin categories.

To convert a new score to a log likelihood ratio, one can search the training scores to find the closest score value to the new score value and then map the new score to the log likelihood ratio corresponding to that training score. One could also interpolate between log-likelihood-ratio values corresponding to the closest lower-value training score and the closest higher-value training score.

The pool-adjacent-violators procedure is equivalent to a histogram model with adaptive bin widths.[11] The bin widths are identical for the same-origin histogram and the different-origin histogram, and are determined by the pool-adjacent-violators algorithm: The width of a bin spans all the scores (same-and different-origin) in a pool. If each histograms is scaled so that its total area sums to 1, then the relative height of corresponding bins (one from the same-origin histogram and one from the different-origin histogram) will be the ratio of two likelihoods. All score values falling within the range of a single bin are mapped to the same likelihood-ratio value.

A pool-adjacent-violators model was trained using the same-origin and different-origin similarity-and-typicality training scores (making use of the `opt_loglr` function in the FOCAL TOOLKIT; Brümmer, 2005). The model was then used to transform the suspect-and-offender scores to likelihood ratios (mapping to the log-likelihood-ratio value corresponding to the closest training-score value). The score-to-log-likelihood-ratio mapping function for the first sample set drawn from the simulated population is shown in Fig 9b (dotted line). The likelihood-ratio results are shown in Fig. 16. The median RMS error value was 0.072, see Fig. 3 for a visual comparison of RMS error values across different approaches. This approach using similarity-and-typicality scores and converting the scores to likelihood ratios using the pool-adjacent-violators algorithm produces results which may be considered close to the true likelihood ratios – they are the closest results obtained form any scoring method considered so far, closer than the results from the kernel-density model.

---

[11] My thanks to Peter Vergeer for pointing this out to me.

Since the non-parametric pool-adjacent-violators approach can be applied irrespective of the distribution of the scores, it can be used to compare all the different score types considered here. Any resulting differences can then be attributed only to differences in the type of score rather than to differences due to the model used for score-to-likelihood-ratio conversion. Fig. 17 provides boxplots of the RMS error values calculated over the 100 sample sets for each of the different types of scores. Note that the overall pattern of relative performance between different score types is the same as that seen in Fig. 3 – the best performance was obtained for the similarity-and-typicality scores.

The pool-adjacent-violators approach should, however, should be used with caution. As a non-parametric procedure, to a greater extent than even the semi-parametric kernel-density procedure, it may be overly dependent on the training data and not generalised well to new data. A particularly important new datum is the suspect-and-offender score. Around certain score values there may be large jumps in the likelihood-ratio value in response to small changes in the score value, and over other ranges of score values there may be an unchanging likelihood-ratio value in response to large changes in score value (a example of the results of this can be seen in Fig. 16a around $x$ values of 2 to 3). Because of these potential problems, a more stable parametric or semi-parametric procedure for converting from scores to likelihood ratios may be preferred to the non-parametric pool-adjacent-violators approach. Ramos & González-Rodríguez (2013), for example, state that they do not propose that the pool-adjacent-violators procedure be used as a method for actually calculating likelihood-ratio values in casework. They use it only as part of a procedure for calculating a minimal log-likelihood-ratio cost ($C_{llr}$) as a measure of system performance.[12] Comparing Figs. 2 and 17 it can be observed that for similarity scores and typicality scores (described below) kernel-density models outperformed pool adjacent violators, and for support-vector-machine scores (described below) logistic regression outperformed pool adjacent violators.[13]

---

[12]I would argue, however, that what matters for the court is the performance of the system which is actually used to calculate the likelihood ratios presented in court. The minimal log-likelihood-ratio cost is therefore not of value for the court unless the pool-adjacent-violators algorithm is an integral part of the system used to calculate the likelihood ratios

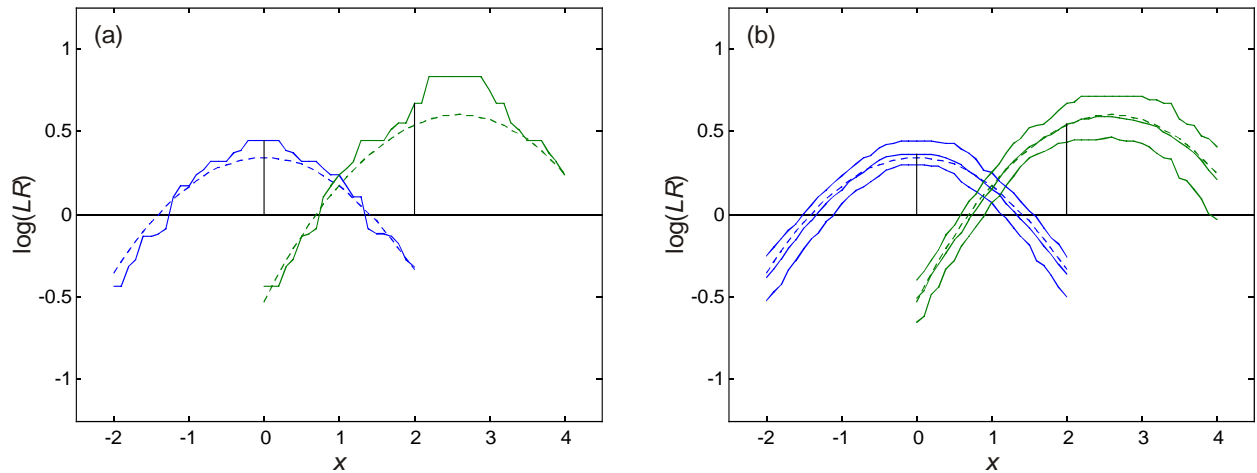

**Figure 16.** Dashed lines: True likelihood-ratio values. (a) Solid lines: Likelihood-ratio values calculated using similarity-and-typicality scores plus pool-adjacent-violators procedure applied to the first set of Monte Carlo sample data. (b) Solid lines: 5th percentile, median, and 95th percentile of likelihood-ratio values calculated using similarity-and-typicality scores plus pool-adjacent-violators procedure applied to 100 sets of Monte Carlo sample data.

---

which are presented in court.

[13]Ali (2014) §3.3 fitted kernel-density, logistic-regression, and pool-adjacent-violator models to samples from several simulated score distributions, calculated likelihood ratios, and compared the results with the ideal values given the specified populations (note that the population distribution was specified for the scores, not for the original features). The parametric logistic-regression model outperformed the semi- and non-parametric models when the specified distributions resembled equal-variance Gaussians. In terms of accuracy, the logistic-regression model also performed best when the amount of training data was small (20 same-origin and 1000 different-origin scores), even for specified distribution which had very different shapes (shapes which I would not expect for the distributions of similarity-and-typicality scores). Smoothing for the kernel-density model was based on a heuristic, which was a function of the sample standard deviation and number of training scores, and results were non-monotonic. Ali preferred the pool-adjacent-violators model over the logistic-regression model because it had better precision, but overall I would have judged that the better accuracy for the logistic regression model outweighed its poorer precision: there were ranges of score values for which the worst results from the logistic-regression model were better than the best results from the pool-adjacent-violators model.

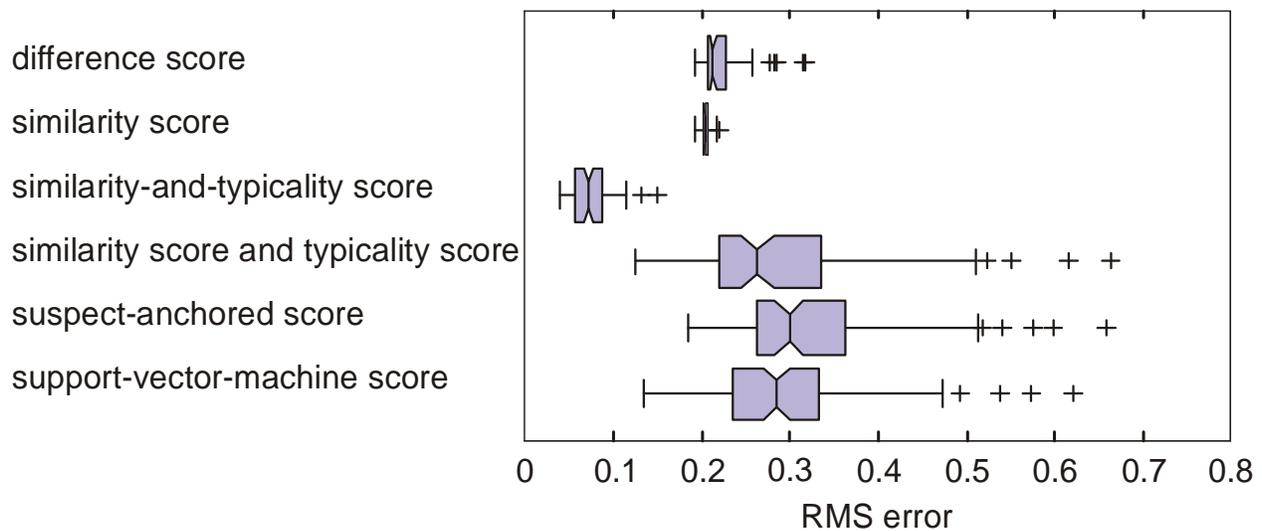

**Figure 17.** RMS error between estimated and true log-likelihood-ratio values calculated over all suspect and offender pairs over 100 sets of samples. Comparisons across different types of scores. Pool adjacent violators used for score-to-likelihood-ratio mapping in every case.

### 5.4 Similarity score (suspect anchored) and typicality score (offender anchored)

Another score-based approach which takes account of both similarity and typicality makes use of suspect-anchored similarity scores in the numerator of the likelihood ratio and offender-anchored typicality scores in the denominator (Drygailo, Alexander, & Meuwly, 2003; Alexander & Drygailo, 2004; Alexander, 2005; González-Rodríguez et al., 2006; Neumann, et al., 2006; Neumann, et al., 2007; Haraksim & Meuwly, 2013, the latter includes a very clear graphical representation of the procedure; see also Hepler et al., 2012, "trace-anchored" approach; Alberink, de Jongh, & Rodríguez, 2014, "asymmetric" approach). There are multiple variants, but for concreteness the procedures for a representative variant are as follows:

1. A generative suspect model is trained using a set of suspect data, the "suspect reference database".

2. The probability density of the suspect model is evaluated at the value of each token in a second set of suspect data, the "suspect control database". The logarithms of these likelihood values constitute a set of similarity scores, $S_{i,k}$ where $i$ indexes one of a series of control database samples and $k$ indexes the suspect model.

3. Generative models are trained using data from a "potential population database" (background database). One model is trained for each source in the potential population database using data from that source.

4. The probability density of the model of each source in the potential population database is evaluated at the value of the offender token. The logarithms of these likelihood values constitute a set of typicality scores, $S_{q,j}$ where $q$ indexes the offender sample and $j$ indexes one of a series of source models from the potential population database.

5. The probability density of the suspect model is evaluated at the value of the offender token. The logarithm of this likelihood value is referred to as the "statistical evidence value", and is our suspect-and-offender score $S_{q,k}$. What $S_{q,k}$ has in common with the similarity scores $S_{i,k}$ is that it is a log likelihood calculated using the suspect model (note that the subscript $k$ is common to both). What $S_{q,k}$ has in common with the typicality scores $S_{q,j}$ is that it is a log likelihood evaluated at the value of the offender token (note that the subscript $q$ is common to both).

6. A generative model is trained on the similarity scores $S_{i,k}$ and another generative model is trained on the typicality scores $S_{q,j}$. The probability density of each model is evaluated at the evidence value $S_{q,k}$, and the likelihood value from the similarity-score model is divided by the likelihood value from the typicality-score model to arrive at a likelihood ratio.

The similarity score and typicality score approach as described above was implemented, calculating similarity scores using a Gaussian model for the suspect model and calculating typicality scores using a separate Gaussian model for each source in the potential population database. The pooled standard deviation was used for all models. A kernel-density model was then used to model the similarity scores and another kernel-density model to model the typicality scores. The bandwidth for the kernel-density models, $\sigma_b = 0.4$, was chosen somewhat arbitrarily on the basis of eyeballing probability-density plots and selecting a value which seemed to give a reasonable degree of smoothing. Note that there was a different similarity model for each suspect and a different typicality model for each offender, hence the likelihoods of each suspect-and-offender score were evaluated given a unique pair of suspect and offender models. Fig. 18a,c,e (left panels) plot the probability-density-function models corresponding to three suspect and offender pairs ($\mu_k = 2$, $x_q = 0$; $\mu_k = 2$, $x_q = 2$; $\mu_k = 2$, $x_q = 4$) and Fig. 18b,d,f (right panels) plot the corresponding score to log likelihood ratio mappings. The suspect-and-offender score values for these pairs ($S_{q,k} = -1.458$, $-0.345$, and $-1.458$ respectively) are the only values at which each of these particular pairs of similarity and typicality models are actually evaluated. These score values are marked as the vertical lines in Fig. 18. Fig. 19 shows the resulting likelihood ratios. The median RMS error value was 0.175, see Fig. 3 for a visual comparison of RMS error values across different approaches. The likelihood-ratio values produced by these similarity score and typicality score procedures are not particularly close to the true likelihood-ratio values – performance is worse than for similarity-and-typicality scores plus logistic-regression calibration.

Thiruvaran et al. (2013) obtained better performance on metrics such as $C_{llr}$ for forensic-voice-comparison systems using similarity-and-typicality scores plus logistic-regression calibration than for parallel systems using similarity scores for the numerator and typicality scores for the denominator plus generative models (combinations of kernel density models and single Gaussian models were tested) plus logistic-regression calibration.

The similarity score and typicality score approach also has the disadvantage of requiring an additional suspect sample, the suspect control database. Botti, Alexander, & Drygajlo (2004) described a variant not requiring a suspect control database, in which similarity scores were modelled using within-source data from other sources (i.e., the numerator was non anchored rather than suspect anchored), but they then encountered the same non-monotonicity problem described above for the separate-variance Gaussian model. Ramos-Castro et al. (2006) described a solution designed to work with a limited amount of additional suspect data: a large number of non-anchored same-source scores are derived, then a Gaussian distribution based on these is adapted towards a Gaussian distribution based on the small number of suspect-anchored scores available. The larger the proportion of suspect-anchored scores, the greater the degree of adaptation.

Amount of kernel-density smoothing seemed to be critical for the specific procedures adopted here. This is likely due to the fact that the similarity model in our example is only based on 30 scores corresponding to the 30 tokens in the suspect control database. Note that there is non-monotonicity in the score to log-likelihood-ratio conversion functions (Fig. 18b,d,f), caused primarily by a scarcity of same-origin training scores with a value of around 2. Although it did not affect any of the specified test data employed here, it is of concern with respect to the generalisability of this approach. These problems would likely be ameliorated if the size of the suspect control database were increased, but the amount of suspect data available in a real case is usually limited and we have already doubled the amount used compared to the amount used in each of the other approaches described so far.

Neumann & Saunders (2015) have also criticised the suspect-anchored numerator and offender-anchored denominator approach as not theoretically defensible.

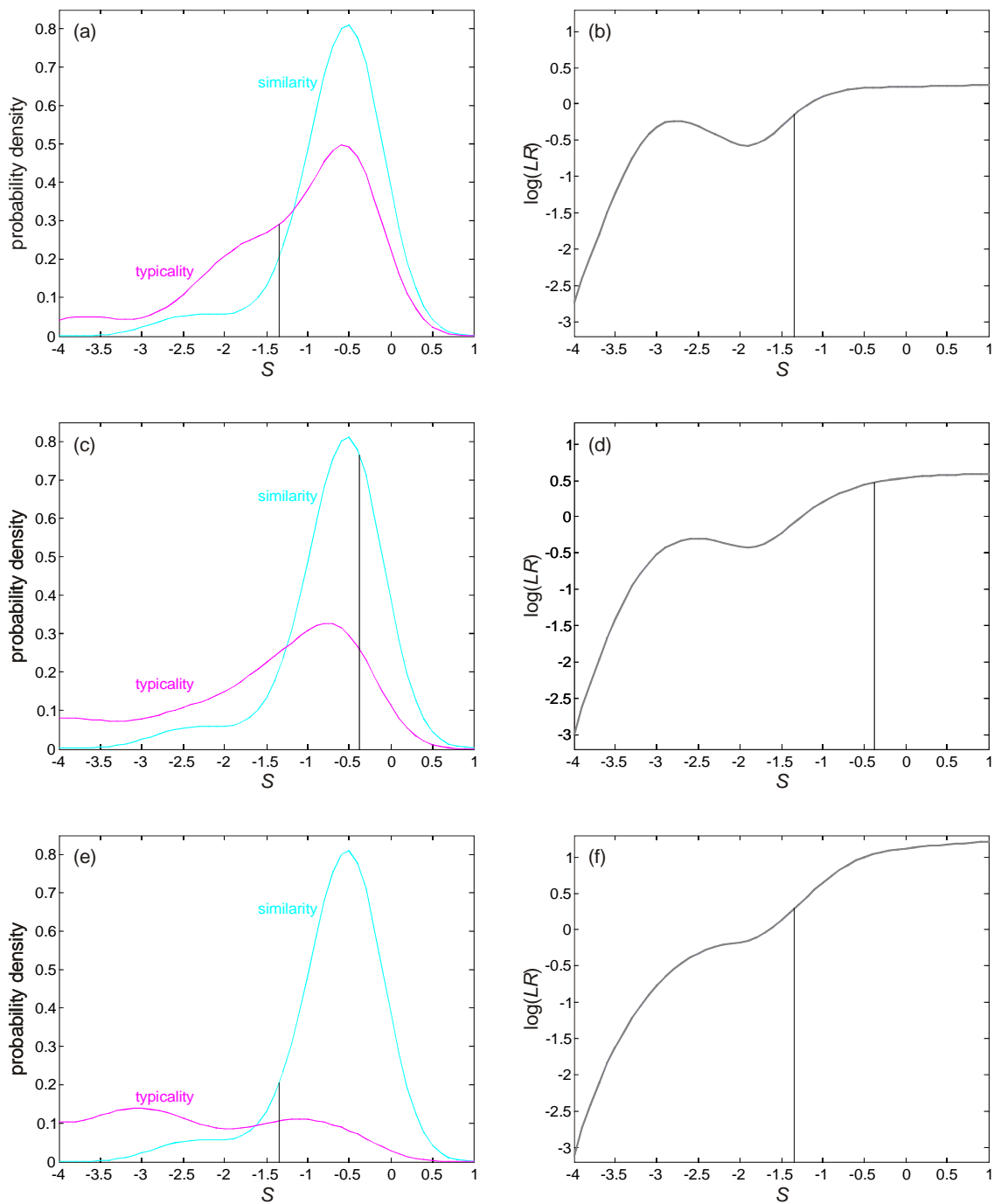

**Figure 18.** Left panels (a,c,e): Modelled distributions of similarity scores and typicality scores (kernel-density

models). Right panels (b,d,f): Score-to-log$_{10}$-likelihood-ratio mapping based on the distributions shown in left panels. Vertical lines shown the suspect-and-offender score values at which these functions are evaluated. Scores are derived from a a suspect mean of 2 in each case and an offender token value of 0 for the top panels (a,b), 2 for the middle panels (c,d), and 4 for the bottom panels (e,f).

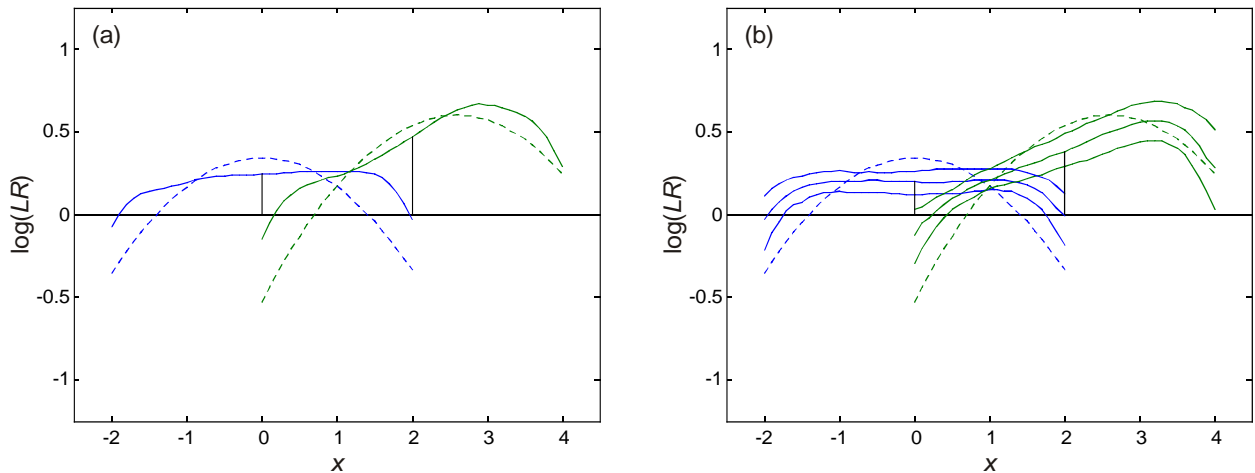

**Figure 19.** Dashed lines: True likelihood-ratio values. (a) Solid lines: Likelihood-ratio values calculated using similarity scores and typicality scores plus kernel density modelling applied to the first set of Monte Carlo sample data. (b) Solid lines: 5th percentile, median, and 95th percentile of likelihood-ratio values calculated using similarity scores and typicality scores plus kernel density modelling applied to 100 sets of Monte Carlo sample data.

## 5.5 Suspect-anchored scores in both numerator and denominator

Variants of the similarity score and typicality score approach have been proposed in which the denominator of the likelihood ratio is anchored on the suspect rather than the offender (see Hepler et al., 2012, "known-source-anchored" approach; Alberink, de Jongh, & Rodríguez, 2014, "fingerprint-anchored" and "finger-anchored" approaches; Ali et al., 2013, "suspect-anchored" approach; see also Ali, 2014). The offender-anchored typicality scores above were calculated by evaluating the likelihood of models of background sources at the value of the offender token. If, instead, suspect-anchored typicality scores were calculated by evaluating the likelihood of the suspect model at the values of tokens in the background database, and the tokens in the suspect reference database had some degree of separation from the offender token in the original feature space, then the distribution of the suspect-anchored typicality scores would differ from that of the offender-anchored typicality scores. This would in turn result in different values for the final likelihood-ratio output.

Conceptually, the use of suspect-anchored scores for calculating the denominator of the likelihood ratio would seem to be problematic in that this is at least partially conditioning the denominator of the likelihood ratio on the suspect. It has long been recognized that the relevant population, and therefore the model for the denominator of the forensic likelihood ratio, should be conditioned on the offender, not the suspect (Robertson & Vignaux, 1995, §3.1; Aitken & Taroni, 2004, ch. 8). If the defence hypothesis is true, as must be assumed for calculating the denominator of the likelihood ratio, information about the suspect does not tell us anything about the offender. For example, knowing the ethnic origin of the suspect provides no information about the ethnic origin of the offender. Thus, specification of the relevant population in this example should not be based on the ethnic origin of the suspect. If there were eyewitnesses who said that the offender appeared to be of a particular ethnic origin, then this information could be used for specifying the relevant population, i.e., people who appear to be of that ethnic origin (note that appearing to witnesses to be of a particular ethnic origin is not the same as actually being of that ethnic origin). If the suspect is not the offender, conditioning the selection of background data on the suspect as opposed to on the offender may lead to differences in the distribution of the background data in the original feature space, which in turn would lead to differences in the distribution of scores calculated using the background data, which in turn would lead to differences in the values of the likelihood-ratio output. Likewise, anchoring typicality scores on the suspect uses information about the suspect in calculating the denominator of the likelihood ratio. In this case information contained in the suspect reference data is used in calculating typicality scores. This may lead to different score distributions than if offender anchoring had been used.

The suspect-anchored approach can be conceptualised in a different way: The suspect-and-offender score quantifies the degree of similarity between the suspect model and the offender token, and this is evaluated relative to scores which quantify the degree of similarity between the suspect model and tokens known to be from the suspect versus scores which quantify the degree of similarity between the suspect model and tokens known to be from other sources. Both numerator and denominator can therefore be conceptualised as modelling the distribution of similarity-to-suspect scores rather than the scores for the numerator being conceptualised as similarity scores and those for the denominator as typicality scores. We have already illustrated that non-anchored similarity-only scores for both the numerator and denominator are not suitable for calculating forensic likelihood ratios. The question now is whether using suspect-anchored similarity scores for both the numerator and denominator leads to appropriate results.

Suspect models consisting of pooled-standard deviation Gaussians with means at the specified $\mu_k$ values were evaluated at the values of every token from every source in the background data.[14] The logarithm of these likelihoods were used as scores for the denominators. These scores were modelled using a kernel-density model with a bandwidth, $\sigma_b = 0.4$. The numerator and denominator distributions and score-to-log-likelihood-ratio mappings were similar to those shown in Fig. 18c and 18d. The distributions of the scores for the numerators were identical, and the distributions of the scores for the denominators were somewhat smoother. Fig. 20 shows the results of the likelihood ratio calculations. The median RMS error value was 0.204, see Fig. 3 for a visual comparison of RMS error values across different approaches. Performance was worse than for the similarity score and typicality score offender-anchored-denominator approach. Something about the atypicality of the  suspect seems to be captured – in contrast to the results of the similarity-only score system shown in Fig. 8, the likelihood-ratio values associated with the rightmost suspect in Fig. 20 are higher than those associated with the leftmost suspect. The atypicality of the offender tokens, however, does not appear to have been captured – as with the results of the similarity-only score system shown in Fig. 8, the likelihood-ratio values in Fig. 20 are symmetric about the rightmost suspect's mean. Empirically, the suspect-anchored approach does not appear to be appropriate for calculating forensic likelihood ratios.

Neumann & Saunders (2015) have also criticised the suspect-anchored numerator and denominator approach as not theoretically defensible.

---

[14]Another variant was also tested. In this variant a suspect model for every source in the background was evaluated at the values of every token in the suspect control database. Results were very close to those reported for the variant described above.

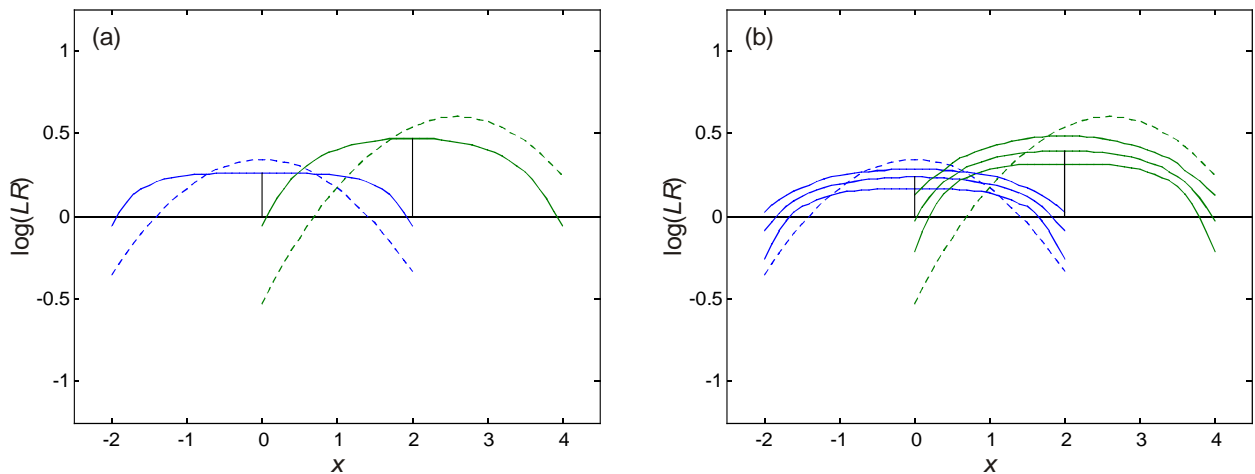

**Figure 20.** Dashed lines: True likelihood-ratio values. (a) Solid lines: Likelihood-ratio values calculated using suspect-anchored scores in both numerator and denominator (second variant) plus kernel density modelling applied to the first set of Monte Carlo sample data. (b) Solid lines: 5th percentile, median, and 95th percentile of likelihood-ratio values calculated using suspect-anchored scores in both numerator and denominator (second variant) plus kernel density modelling applied to 100 sets of Monte Carlo sample data.

## 5.6 Support-vector-machine scores

Another approach which has been applied to the calculation of forensic likelihood ratios is the use of a support vector machine followed by logistic-regression calibration (Platt, 2000; González-Rodríguez et al., 2007; Abraham et al., 2006). See Hastie, Tibshirani, & Friedman (2009) pp. 417–434 for a detailed description of support vector machines. Briefly: A support vector machine is a discriminative method. Typically a support vector machine expands the raw feature space using basis functions (kernels) and then fits a linear discriminator in the expanded space. In general the linear discriminator is a separating hyperplane. With a hard margin, the hyperplane is placed so that all training data from one category are on one side of the hyperplane and all training data from the other category are on the other side, and the perpendicular distance between the hyperplane and the nearest training points from each category is maximised. The nearest points are the support vectors and their distance from the hyperplane is the margin width. With a hard margin the support vectors lie on the margin boundary. In a two-dimensional space, three support vectors are needed to determine the location of a separating line, two support vectors from one category and one from the other. The line is drawn so that it is equidistant from each of the three support vectors, each point is the margin distance away from the line, two in one direction and one in the other direction. A three-dimensional space requires four support vectors to determine the location of a separating plane, etc. With a soft margin, support vectors can be closer to the hyperplane than the margin width and can even be on the wrong side of the separating hyperplane. A larger number of support vectors are involved, but their average distance from their respective margin boundary is minimised (distance measured in the direction perpendicular away from the respective margin boundary towards the hyperplane). Training points on the correct side of the hyperplane falling outside the margin do not contribute (or have minimal contribution) to determining the location of the hyperplane.

Hard classification is performed by projecting a test datum into the expanded space and then classifying according to which side of the hyperplane it falls on. A soft classification can be achieved by measuring, in the expanded space, the distance of the test datum from the separating hyperplane in the direction perpendicular to the hyperplane. Distances on one side of the plane are positive and distances on the other side negative. This distance we will refer to as a support-vector-machine score.

Support vector machines do not model the distribution of the background data (the sample of the population) and typically only use a small portion of the background data as support vectors. Because of this it may be that, for the purpose of calculating forensically interpretable likelihood ratios, they do not adequately account for typicality with respect to the relevant population. We proceed to examine empirical results.

Support vector machines were trained using 30 sample tokens from the suspect and 3000 tokens from the background sample, i.e., 30 sample tokens from each of 100 sampled sources. The 30 suspect tokens were generated on the basis of a mean of 0 for one suspect and 2 for the other suspect and a standard deviation of 1 in both cases. The generated suspect tokens were used as is, and were not recentred such that their sample mean would be exactly 0 or 2. The choice of basis function may be a critical factor in the success of this approach if the aim is to derive interpretable likelihood ratios. A Gaussian radial basis function with a standard deviation of 1.5 was used. This is akin to fitting a Gaussian kernel density model and therefore seemed a good candidate. Support vector machines were trained using MATLAB's `svmtrain` function with default values other than those specified above for the basis function.

Scores were derived for the same 30 suspect tokens and 3000 background tokens as had been used to train the support vector machine. These scores were then used to train a logistic regression model. A score for each offender token was obtained from the support vector machine and converted to a likelihood ratio using the coefficient values from the logistic regression model. The results are shown in Fig. 21. The median RMS error value were 0.149, see Fig. 3 for a visual comparison of RMS error values across different approaches. Performance was not as good as that obtained using similarity-and-typicality scores plus logistic-regression calibration.

Performance turned out to be sensitive to the value chosen for the standard deviation of the Gaussian radial basis function, particularly performance related to the rightmost suspect and highest offender-token values. Fig. 22 shows the results of using a standard deviation of 1 and of 3. No value was found that clearly ameliorated the problem across the 100 sample sets, especially if one were to take into account expected performance on more atypical suspect and offender combinations than those tested here. The problem is likely related to the fact that the model was trained on only 30 suspect tokens, but this is the same number as was used in each of the other approaches described above (except for the similarity score and typicality score approach, and the suspect-anchored approach, which both had an additional 30 tokens as a suspect control database). As noted above, the amount of suspect data available in a real case is usually limited.

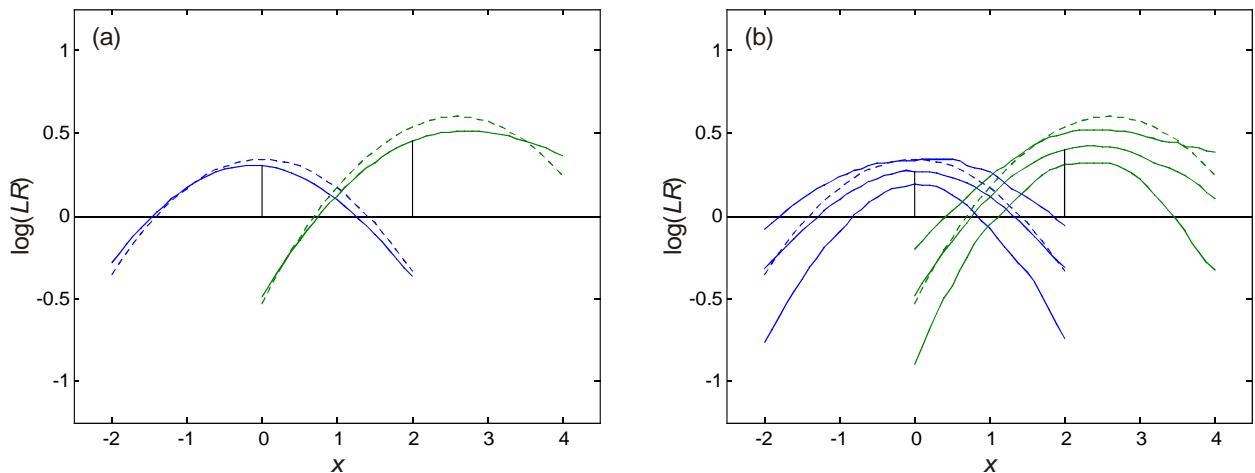

**Figure 21.** Dashed lines: True likelihood-ratio values. (a) Solid lines: Likelihood-ratio values calculated using support-vector-machine scores plus logistic-regression calibration applied to the first set of Monte Carlo sample data.

(b) Solid lines: 5th percentile, median, and 95th percentile of likelihood-ratio values calculated using support-vector-machine scores plus logistic-regression calibration applied to 100 sets of Monte Carlo sample data.

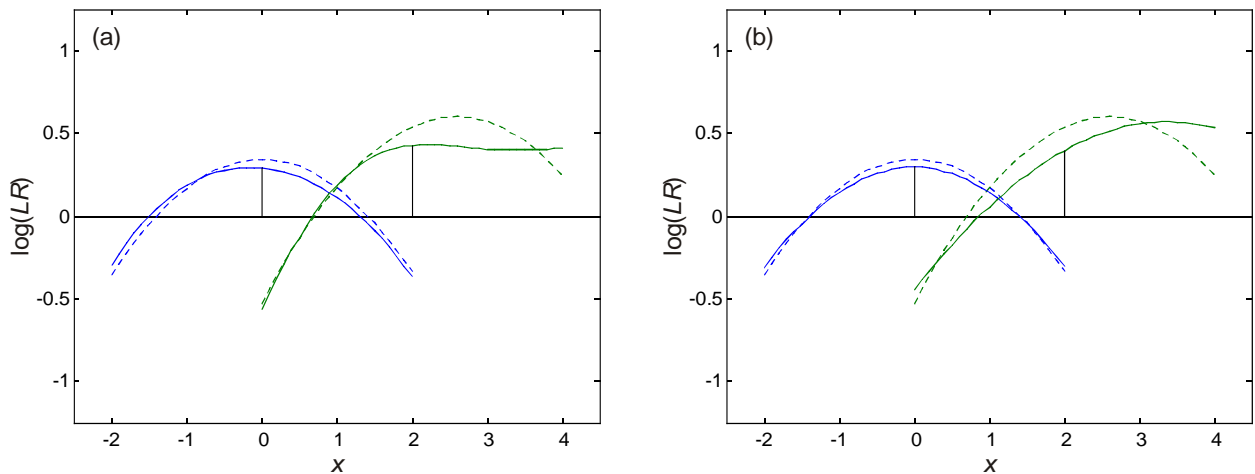

**Figure 22.** Dashed lines: True likelihood-ratio values. Solid lines: Likelihood-ratio values calculated using support-vector-machine scores plus logistic-regression calibration applied to the first set of Monte Carlo sample data. (a) Radial basis function standard deviation of 1. (b) Radial basis function standard deviation of 3.

## 6  CONCLUSION

The present paper has explored a number of score-based approaches for calculating forensic likelihood ratios, using Monte Carlo simulation to compare their output to true likelihood-ratio values derived from a simulated population.

Some approaches should be ruled out on theoretical grounds. Approaches based on difference-only scores or similarity-only scores do not take account of typicality with respect to the relevant population, and do not, therefore, produce forensically interpretable likelihood ratios.

Scores which take account of both similarity and typicality can be used to calculate forensically interpretable likelihood ratios, but an appropriate score-to-likelihood-ratio conversion procedure must be used. Such a procedure should be a model which either directly calculates the ratio of two likelihoods or which is analogous to such a model. It should also produce a monotonic mapping between score values and likelihood-ratio values, at least over the range of score values which it is ever expected to encounter. It should ideally also be robust.

Logistic-regression calibration applied to similarity-and-typicality scores (non-anchored scores for both the numerator and denominator of the likelihood ratio) fulfilled all of the desiderata listed above. Logistic regression was found to produce likelihood-ratio values which may be considered reasonably close to the true likelihood-ratio values. Output substantially closer to true likelihood-ratio values were obtained using kernel-density models applied to similarity-and-typicality scores, but this approach requires care in selecting parameter values related to degree of kernel smoothing and care to ensure that the models are only ever presented with suspect-and-offender data within their proven operating range. The pool-adjacent-violators procedure applied to similarity-and-typicality scores resulted in likelihood-ratio values which were yet closer to the true likelihood-ratio values, but this procedure is susceptible to overtraining on the training scores and so should also be used with caution.

Unlike logistic regression, equal-variance Gaussian models applied to similarity-and-typicality scores are not robust to violations of modelling assumptions, and the test data violated the assumptions. Separate-variance Gaussian models applied to similarity-and-typicality scores do not result in monotonic score-to-likelihood-ratio mappings.

An approach using kernel-density models of similarity scores (suspect-anchored scores) to calculate the numerator of the likelihood ratio and kernel-density models of typicality scores (offender-anchored scores) to calculate the denominator did not produce likelihood-ratio values which were particularly close to the true likelihood-ratio values. The same was true for an approach using support-vector-machine scores plus logistic-regression calibration. Both were outperformed by similarity-and-typicality scores plus logistic-regression calibration, and they appeared to be sensitive to the choice of parameter settings and to any peculiarities of the limited amount of suspect data available (limited suspect data is problem which is common in casework).

An approach in which suspect-anchored scores were used to calculate both the numerator and denominator of the likelihood did not capture the degree of atypicality of offender tokens. Such an approach is also be theoretically problematic in that it conditions the denominator of the likelihood ratio on suspect information.

Overall, the best score-based approaches for the particular simulated population tested here appear to be to use similarity-and-typicality scores, and to convert the scores to likelihood ratios using logistic-regression calibration, kernel-density models, or the pool-adjacent-violators algorithm. Logistic regression may be the best overall solution because of its greater robustness.

One should be cautious of generalising to other contexts a choice of one approach over another on the basis of the empirical results presented above. Only one simulated population was tested, and empirical results may differ for more complex multivariate and multimodal population distributions to which these models are normally applied. The exact nature of the scores derived and the manner in which models are applied in other contexts may also differ. The key message instead is the recommendation that for any proposed procedure for calculating forensically interpretable likelihood ratios, whenever possible, Monte Carlo simulation should be used to compare the output of the procedure against true likelihood-ratio values from a simulated population. Ideally, the simulated test data should reflect as best as possible the conditions of the real data to which the procedure will be applied. Monte Carlo testing has limitations because the true distribution of real population data may be more complex than that of the simulated population, or simply differ from that of the simulated population. Monte Carlo simulation can, however, be used to ask how well a procedure might work if the structure of the population data actually matched the structure assumed by the model being tested. The model's assumed data structure would be a best guess at what the structure of a real population might be. The initial parameter values for the simulated population could be derived by fitting a model to actual data. If the model does not perform well with simulated data which match the data structure it assumes, then one would not expect it to perform well with real population data which may have a more complex underlying structure. A procedure which was found to work well using simulated populations would not be guaranteed to work well on real population data, but such an approach to method validation would at least allow us to discount proposed procedures which are found to perform poorly even under these favourable conditions.

## APPENDIX A - Multivariate Multimodal Simulation

The simulations described below are designed to explore whether an approach based on similarity-and-typicality scores outperforms an approach based on difference-only or similarity-only scores for a more complex simulated populations than that described in §2. Specifically, the simulated populations below are multivariate and multimodal. Populations were generated in a manner analogous to that described in §2. The population distribution had an initial centroid at the origin of a multivariate space. Rather than using a single Gaussian to generate the between-source distribution, multiple Gaussians were used. The mean vector for each between-source Gaussian was generated pseudo-randomly from a uniform distribution within a predefined range in each dimension. Each of the between-source Gaussians had an initial covariance matrix which was diagonal and which had the same variance for each dimension. For each Gaussian, tokens were pseudo-randomly generated (the number of tokens generated per Gaussian was equal to the number of dimensions multiplied by 6), and these tokens were used to calculate a new mean vector and covariance matrix. Given the new mean vector and covariance matrix values for the between-source Gaussians, 1000 points were pseudo-randomly generated per Gaussian. Each of these points was then the initial centroid for a source. Each within-source distribution also consisted of multiple Gaussians, the mean vectors of which were pseudo-randomly generated from a uniform distribution within a predefined range in each dimension (a smaller range than was used for the between-source distribution). The initial covariance matrix for each within-source Gaussian was diagonal and had the same variance for each dimension (a smaller variance than was used for the between-source distribution). Again, for each Gaussian, tokens were pseudo-randomly generated (the number of tokens generated per Gaussian was equal to the number of dimensions multiplied by 6), and these tokens were used to calculate a new mean vector and covariance matrix. These mean vectors and covariance matrices were saved as the parameters for the population.

A number of additional sources were pseudo-randomly generated to use as suspect sources, and their mean vectors and covariance matrices saved. Offender values were selected on diagonal lines running though the feature space. Each diagonal straddled the space corresponding to a specified suspect model. The start point on each dimension was selected as follows: For each Gaussian specified for a given suspect model the point 1 standard deviation below the mean was calculated, then the minimum value across all the Gaussians for that suspect was selected. Likewise, the end point on each diagonal was the maximum of 1 standard deviation above the mean of each Gaussian for that suspect. 12 offender points were equally spaced along the multidimensional diagonal from the start point to the end point.

True likelihood-ratio values were calculated for the suspect and offender combinations using the saved parameter values for the suspect distributions and population distribution. Each within-source Gaussian was given equal weight. Monte Carlo samples were generated by pseudo-randomly selecting a source Gaussian (equal probability of selecting each source), then for that source pseudo-randomly selecting a withing-source Gaussian (equal probability of selecting each of the Gaussians within that source), and generating a token given the saved mean vector and covariance matrix for that within-source Gaussian. The process was repeated to generate the desired number of tokens from the desired number of sources. The tokens were then used to train models, and the likelihood-ratio values resulting from these models were compared with the true likelihood ratio values.

Exploration was conducted using Monte Carlo simulations with different numbers of dimensions, different numbers of between- and within-source Gaussians, and different spacing between mean vectors (ranges of uniform distributions) for between- and within source, and different magnitudes for valances for between- and within source. Methods tested were direct calculation of likelihood ratios, difference-only scores plus pool adjacent violators, similarity-only scores plus pool adjacent violators, and similarity-and-typicality scores plus pool adjacent violators. Pool adjacent violators was selected because it could be applied to both types of score without concern for the details of the distributions of the scores, and any difference in performance could then be credited to a difference in the type of score used and not to a potential difference in the choice and fitting of score-to-likelihood-ratio-mapping functions. Models used for calculating the numerator and the denominator in the direct method, and for deriving the scores in the similarity-only and the similarity-and-typicality scoring methods, included a single multivariate Gaussian using least-squares fit, and Gaussian mixture models fitted using expectation maximisation. The models were fit using MATLAB's `gmdistribution.fit` function, using a regularisation value of 0.001, and using 4 different random seeds and keeping the result with the largest likelihood for the training data. The difference-only scores were calculated as the mean of the Euclidian distances between a single offender token and each of the tokens of a suspect. To reduce computing time, rather than using cross-validation, same-origin training scores were calculated by building a model using one set of values generated for a source and evaluating the likelihood of that model a values from a second set of values generated for the same source. Likewise, different-origin training scores were calculated by comparing a model trained on sample data from one source with sample data from a second source, where the second source came from a second set of sources selected pseudo-randomly form the population. The first and second set of sources did not overlap, and each source was used only once in calculating a set of different-origin scores.

One observation from the exploration is that with a relatively small amount for training data the best model was usually a direct model using a single Gaussian, unless there was a wide separation between within-source Gaussians.[15] If the Gaussians used to generate the sample data are relatively close together, especially if there is a relatively large number of Gaussians, then they sum of the sample data generated from the multiple Gaussians approximates the distribution expected from a single Gaussian. Only if the Gaussians are relatively widely spaced will the sample data be sufficiently multimodal that a single Gaussian model will be a poor fit to the data compared to a Gaussian mixture model. A second observation is that if the amount of training data was increased then a direct method based on Gaussian mixture models would outperform a scoring method even if the within-source Gaussians were widely separated. An insight from this is that one should consider fitting a model which directly calculates likelihood ratios rather than immediately assuming that a score-based method will provide the best results. If there is sufficient data, then one may be able to obtain a good fit for a direct method requiring estimates for a relatively large number of parameter values, and if there is relatively little data a potentially high bias but low variance direct-method model may outperform a scoring method.

Both the similarity-only scores and the similarity-and-typicality scores always outperformed the distance-only scores by a large margin, and the results for the difference-only scores will not be discussed further.

Since we are interested in situations in which direct methods do not outperform scoring methods, we added a constraint to the Monte Carlos simulations such that there was a minimum separation between the within-source mean vectors (a single set of offset values was generated, then for each source those offset values were added to the source centroid to generate the initial mean vectors for that source), and we also used a relatively small number of sample tokens per source. Once the conditions were met such that a direct method did not outperform the best scoring method, we observed that averaged across multiple Monte Carlo sample sets the method based on similarity-and-typicality scores outperformed the method based on similarity-only scores (see for example Fig. A1), but the degree by which similarity-and-typicality scores outperformed similarity-only scores lessened as the number of dimensions or number of Gaussians increased and the amount of training data was unchanged. Eventually, the similarity-only scores slightly outperformed the similarity-plus-typicality scores (see for example Fig. A2); however, at this point the performance of the best performing method was very poor, suggesting that none of the tested methods would be appropriate for a large number of dimensions and modes and a small amount of training data. One should remember not only to ask which method gives the best performance, but also whether any method gives sufficiently good performance. For multidimensional multimodal data and relatively small amounts of training data, scoring methods may outperform direct methods; however, performance of a score-based method may be relatively poor if the number of dimensions and number of modes is high and the amount of training data is small. Scoring methods may outperform direct methods when the latter are afflicted by the curse of dimensionality, but scoring methods themselves are not immune and are also eventually afflicted by the curse of dimensionality.

---

[15]Since the denominator (typicality) models were trained using data pooled from all sampled sources, the amount of data available to train such models was larger, hence the distribution of the between-source Gaussians was not as critical.

Figs. A1 and A2 give examples of boxplots of RMS errors based on a simulated population with 4 between-source Gaussians and 1000 sources per between-source Gaussian (a total of 4000 sources). The uniform distribution from which the initial values of the between-source mean vectors were generated had a range of −4 to +4 on each dimension, and the initial variances were 1. The uniform distribution from which the initial values of the within-source mean vectors were generated had a range of −2 to +2 on each dimension with the constraint that the Euclidian distance between each Gaussian and its nearest neighbour not be less than 2, and the initial variances were 0.25 (the initial mean vectors were a minimum of 4 standard deviations apart). Each sample set consisted of 100 sources and 30 tokens per source. The number of suspects was 12 per between-source Gaussian (a total of 48). The results in Fig. A1 come from a 2-dimensional distribution with 3 within-source Gaussians per source. The similarity scores were calculated using a Gaussian mixture model consisting of 3 Gaussians. The results in Fig. A2 come from a 4-dimensional distribution with 5 within-source Gaussians per source. The similarity scores were calculated using a Gaussian mixture model consisting of 5 Gaussians. In both cases the denominator for the similarity-plus-typicality scores was calculated using a Gaussian mixture model consisting of 4 Gaussians trained using data pooled across all sources. In Fig. A1, the method based on similarity-and-typicality scores clearly outperforms that based on similarity-only scores, and the size of the RMS errors may be considered acceptable – the median value is 0.57. In Fig. A2, the method based on similarity-only scores appears to slightly outperform that based on similarity-and-typicality scores, but the size of the RMS errors may be considered unacceptable – the median value for similarity-only scores is 3.72.

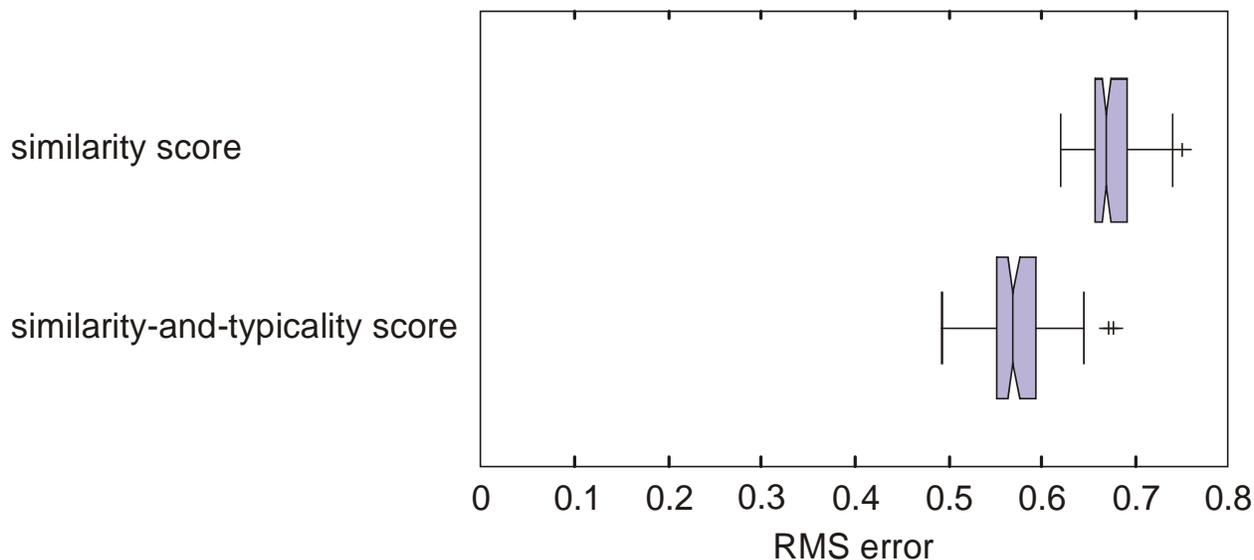

**Figure A1.** RMS error between estimated and true log-likelihood-ratio values calculated over all suspect and offender pairs over 100 sets of samples from a 2-dimensional distribution generated using 4 between-source Gaussians and 3 within-source Gaussians. Comparison between methods based on similarity-only scores and similarity-and-typicality scores. Pool adjacent violators used for score-to-likelihood-ratio mapping in both cases.

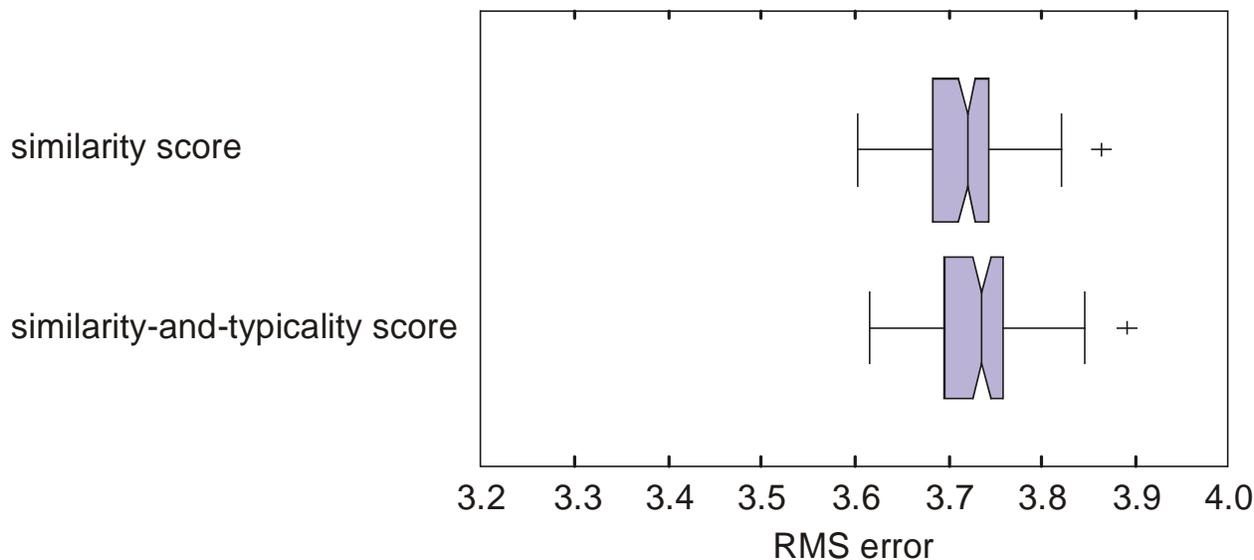

**Figure A2.** RMS error between estimated and true log-likelihood-ratio values calculated over all suspect and offender pairs over 100 sets of samples from a 4-dimensional distribution generated using 4 between-source Gaussians and 5 within-source Gaussians. Comparison between methods based on similarity-only scores and similarity-and-typicality scores. Pool adjacent violators used for score-to-likelihood-ratio mapping in both cases. Note the scale on the ordinate and that it does not begin at 0.

**REFERENCES**

Abraham J., Champod C., Lennard C., Roux C. (2013). Modern statistical models for forensic fingerprint examinations: A critical review. *Forensic Science International, 232*, 131–150. doi:10.1016/j.forsciint.2013.07.005

Abraham J., Kwan P., Champod C., Lennard C., Roux C. (2012). An AFIS candidate list centric fingerprint likelihood ratio model based on morphometric and spatial analyses (MSA), in J. Yang (Ed.), *New Trends and Developments in Biometrics*. InTech, Rijeka, Croatia, 2012, pp. 221–250. doi:10.5772/51184

Aitken C., Roberts P., Jackson G. (2010). *Fundamentals of Probability and Statistical Evidence in Criminal Proceedings: Guidance for Judges, Lawyers, Forensic Scientists and Expert Witnesses*. Practitioner Guide No 1, Royal Statistical Society's Working Group on Statistics and the Law. http://www.rss.org.uk/site/cms/contentviewarticle.asp?article=1132

Aitken C.G.G., Taroni F. (2004) *Statistics and Evaluation of Evidence for Forensic Scientists*, 2nd ed. Wiley, Chichester, UK, 2004.

Alberink I., de Jongh A., Rodríguez C. (2014). Fingermark evidence evaluation based on automated fingerprint identification system matching scores: The effect of different types of conditioning on likelihood ratios. *Journal of Forensic Sciences, 59*, 70–81. doi:10.1111/1556-4029.12105

Alexander A. (2005). *Forensic automatic speaker recognition using bayesian interpretation and statistical compensation for mismatched conditions*. PhD dissertation, Swiss Federal Institute of Technology in Lausanne, Switzerland.

Alexander A., Drygajlo A. (2004). Scoring and direct methods for the interpretation of evidence in forensic speaker recognition. *Proceedings of Interspeech 2004, Jeju, Korea*, pp. 2397–2400.

Ali T. (2014). *Biometric score calibration for forensic face recognition*. PhD dissertation. University of Twente, The Netherlands.

Ali T., Spreeuwers L., Veldhuis R. (2012). A review of calibration methods for biometric systems in forensic applications. *Proceedings of the 33rd WIC Symposium on Information Theory in the Benelux and The 2rd Joint WIC/IEEE Symposium on Information Theory and Signal Processing in the Benelux, Boekelo, The Netherlands*, 2012, pp. 126–133.

Ali T., Spreeuwers L., Veldhuis R., Meuwly D.(2013). Effect of calibration data on forensic

likelihood ratio from a face recognition system. *Proceedings of IEEE Sixth International Conference on Theory, Applications and Systems (BTAS)*. doi:10.1109/BTAS.2013.6712709

Bååth R. (2014). Bayesian First Aid, R software package. https://github.com/rasmusab/bayesian_first_aid [see also http://www.r-bloggers.com/bayesian-first-aid-one-sample-and-paired-samples-t-test/] last accessed 11 May 2014.

Balding D.J. (2005). *Weight-of-evidence for Forensic DNA Profiles*. Wiley, Chichester, UK.

Botti F., AlexanderA., Drygajlo A. (2004). An interpretation framework for the evaluation of evidence in forensic automatic speaker recognition with limited suspect data. *Proceedings of Odyssey 2004 The Speaker and Language Recognition Workshop, Toledo, Spain*.

Bowman A.W., Azzalini A. (1997). *Applied Smoothing Techniques for Data Analysis: the Kernel Density Approach with S-plus Illustrations*. Oxford University Press, New York.

Brümmer N. (2005). FoCal Toolbox: Tools for fusion and calibration of automatic speaker detection systems. http://niko.brummer.googlepages.com/focal/ last accessed 11 May 2014.

Brümmer N., du Preez J. (2006). Application independent evaluation of speaker detection. *Computer Speech and Language, 20*, 230–275. doi:10.1016/j.csl.2005.08.001

Brümmer N., Swart A., van Leeuwen D. (2014). A comparison of linear and non-linear calibrations for speaker recognition. *Proceedings of Odyssey 2014 The Speaker and Language Recognition Workshop, Joensuu, Finland*, pp. 14–18.

Drygajlo A., Meuwly D., Alexander A. (2003). Statistical methods and Bayesian interpretation of evidence in forensic automatic speaker recognition. *Proceedings of Eurospeech 2003, Geneva, Switzerland*, pp. 689–692.

González-Rodríguez J., Drygajlo A., Ramos-Castro D., Garcia-Gomar M., Ortega-Garcia J. (2006). Robust estimation, interpretation and assessment of likelihood ratios in forensic speaker recognition. *Computer Speech and Language, 20*, 331–355. doi:10.1016/j.csl.2005.08.005

González-Rodríguez J., Rose P., Ramos D., Toledano D.T., Ortega-García J. (2007). Emulating DNA: Rigorous quantification of evidential weight in transparent and testable forensic speaker recognition. *IEEE Transactions on Audio, Speech, and Language Processing, 15*, 2104–2115. doi:10.1109/TASL.2007.902747

Haraksim R., Meuwly D. (2013). Influence of the datasets size on the stability of the LR in the lower region of the within source distribution. *Proceedings of Biometric Technologies in Forensic Science, BTFS 2013, Nijmegen, The Netherlands*, pp.34–38. http://www.ru.nl/clst/btfs/proceedings/

Hastie T., Tibshirani R., Friedman J. (2009). *The Elements of Statistical Learning: Data Mining, Inference, and Prediction*, 2nd ed. Springer, New York.

Hepler A.B., Saunders C.P., Davis L.J., Buscaglia J. (2012). Score-based likelihood ratios for handwriting evidence. *Forensic Science International 219*, 129–140. doi:10.1016/j.forsciint.2011.12.009

Hosmer D.W., Lemeshow S. (2000). *Applied Logistic Regression*, 2nd ed. Wiley, New York.

Kruschke J.K. (2013). Bayesian estimation supersedes the *t* test. *Journal of Experimental Psychology: General, 142*, 573–603. doi:10.1037/a0029146

Mandasari M.I., Günther M., Wallace R., Saeidi R., Marcel S., van Leeuwen D.A. (2014). Score calibration in face recognition. IET Biometrics. doi:10.1049/iet-bmt.2013.0066

MathWorks Inc. (2013). Matlab, software release 2013b, MathWorks Inc., Nantick, MA.

Morrison G.S. (2009). Forensic voice comparison and the paradigm shift. *Science & Justice, 49*, 298–308. doi:10.1016/j.scijus.2009.09.002

Morrison G.S. (2010). Forensic voice comparison, in I. Freckelton, H. Selby (Eds.), *Expert Evidence*. Thomson Reuters, Sydney, Australia, ch. 99. http://expert-evidence.forensic-voice-comparison.net/

Morrison G.S. (2012). Measuring the validity and reliability of forensic likelihood-ratio systems. *Science & Justice, 51*, 91–98. doi:10.1016/j.scijus.2011.03.002

Morrison G.S. (2013). Tutorial on logistic-regression calibration and fusion: Converting a score to a likelihood ratio. *Australian Journal of Forensic Sciences*, 45, 173–197. doi:10.1080/00450618.2012.733025

Morrison G.S., Stoel R.D. (2014). Forensic strength of evidence statements should preferably be likelihood ratios calculated using relevant data, quantitative measurements, and statistical models – a response to Lennard (2013) Fingerprint identification: How far have we come? *Australian Journal of Forensic Sciences, 54*, 245–256. doi:10.1080/00450618.2013.833648

Neumann C., Champod C., Puch-Solis R., Egli N., Anthonioz A., Meuwly D., Bromage-Griffiths

A. (2006). computation of likelihood ratios in fingerprint identification for configurations of three minutiæ. *Journal of forensic sciences, 51*, 1255–1266. doi:10.1111/j.1556-4029.2006.00266.x

Neumann C., Champod C., Puch-Solis R., Egli N., Anthonioz A., Bromage-Griffiths A. (2007). computation of likelihood ratios in fingerprint identification for configurations of any number of minutiæ. *Journal of forensic sciences, 52*, 54–64. doi:10.1111/j.1556-4029.2006.00327.x

Neumann C., Saunders C. (2015). Commentary on Alberink, et al (2014) Fingermark: The effect of different types of conditioning on likelihood ratios. *Journal of Forensic Sciences, 60*, 252–256. doi:10.1111/1556-4029.12634

Pampel F.C. (2000). *Logistic Regression: A Primer*. Sage, Thousand Oaks, CA.

Platt J.C. (2000). Probabilistic outputs for support vector machines and comparisons to regularized likelihood methods, in A.J. Smola, P. Bartlett, B. Schölkopf, D. Schuurmans (Eds.), *Advances in Large Margin Classifiers*. MIT Press, Cambridge, MA, 2000, pp. 61–74.

R Development Core Team (2014). R: A language and environment for statistical computing, ver. 3.1.0, R Foundation for Statistical Computing, Vienna, Austria. http://www.R-project.org/ last accessed 11 May 2014.

Ramos Castro D. (2007). *Forensic evaluation of the evidence using automatic speaker recognition systems*. PhD dissertation. Autonomous University of Madrid, Spain.

Ramos D., González-Rodríguez J. (2013). Reliable support: Measuring calibration of likelihood ratios. *Forensic Science International, 230*, 156–169. doi:10.1016/j.forsciint.2013.04.014

Ramos-Castro D., González-Rodríguez J., Montero-Asenjo A., Ortega-García J. (2006). Suspect-adapted MAP estimation of within-source distributions in generative likelihood ratio estimation. *Proceedings of the Odyssey 2006 Speaker and Language Recognition Workshop*, IEEE. doi:10.1109/ODYSSEY.2006.248090

Robertson B., Vignaux G.A. (1995). *Interpreting Evidence*. Wiley, Chichester, UK.

Tang Y., Srihari S.N. (2014). Likelihood ratio estimation in forensic identification using similarity and rarity. *Pattern Recognition, 47*, 945–958. doi:10.1016/j.patcog.2013.07.014

Thiruvaran T., Ambikairajah E., Epps J., Enzinger E. (2013). A comparison of single-stage and two-stage modelling approaches for automatic forensic speaker recognition. *Proceedings of The IEEE 8th International Conference on Industrial and Information Systems, ICIIS 2013,*

*Sri Lanka*, pp. 433–438. doi:10.1109/ICIInfS.2013.6732023

van Houten W., Alberink I., Geradts Z. (2011). Implementation of the likelihood ratio framework for camera identification based on sensor noise patterns. *Law, Probability and Risk, 10*, 149–159. doi:10.1093/lpr/mgr006

van Leeuwen D.A., Brümmer N. (2007). An introduction to application-independent evaluation of speaker recognition systems, in C. Müller (Ed.), *Speaker Classification I: Selected Projects*. Springer-Verlag, Heidelberg, Germany, 2007, pp. 330–353.

van Leeuwen D.A., Brümmer N. (2013). The distribution of calibrated likelihood-ratios in speaker recognition. *Proceedings of Biometric Technologies in Forensic Science, BTFS 2013, Nijmegen, The Netherlands*, pp. 24–29. http://www.ru.nl/clst/btfs/proceedings/

Vergeer P., Bolck A., Peschier L.J.C., Berger C.E.H., Hendrikse J.N. (2014). Likelihood ratio methods for forensic comparison of evaporated gasoline residues. *Science & Justice*. doi:10.1016/j.scijus.2014.04.008

Zadrozny B., Elkan C. (2002). Transforming classifier scores into accurate multiclass probability estimates. *Proceedings of the Eighth International Conference on Knowledge Discovery and Data Mining, Edmonton, Alberta, Canada*, pp. 694–699.doi:10.1049/iet-bmt.2013.0066